\documentclass[amssymb,aps,amsmath,floatfix,pre,twocolumn,superscriptaddress,longbibliography]{revtex4-2}
\usepackage[latin1]{inputenc}
\usepackage{mathptmx}
\usepackage[T1]{fontenc}
\usepackage{graphicx}
\usepackage{amsmath,amssymb,amsfonts, MnSymbol}
\usepackage{mathrsfs}
\usepackage{bm}
\usepackage{hyperref}
\usepackage{bbold}
\usepackage{xcolor}
\usepackage{tikz}
\usepackage{overpic}
\newcommand{\algn}[1]{\begin{align} #1 \end{align}}
\newcommand{\sbeqs}[1]{\begin{subequations} #1 \end{subequations}}
\newcommand{\pmat}[1]{\begin{pmatrix} #1 \end{pmatrix}}
\newcommand{\css}[1]{\begin{cases} #1 \end{cases}}

\newcommand{\ve}[1]{\boldsymbol{#1}}
\newcommand{\nn}{\nonumber}
\newcommand{\ee}{\ensuremath{\text{e}}}
\newcommand{\ed}{\ensuremath{\text{d}}}
\newcommand{\dd}[1]{\ensuremath{\tfrac{\text{d}}{\text{d} #1}}}

\newcommand{\mc}[1]{\ensuremath{\mathcal{#1}}}
\newcommand{\ms}[1]{\ensuremath{\mathscr{#1}}}
\newcommand{\mbb}[1]{\ensuremath{\mathbb{#1}}}
\newcommand{\kb}{\ensuremath{k_\text{B}}}
\newcommand{\eqnlab}[1]{\label{eq:#1}}
\newcommand{\seclab}[1]{\label{sec:#1}}
\newcommand{\figlab}[1]{\label{fig:#1}}
\newcommand{\eqnref}[1]{\eqref{eq:#1}}
\newcommand{\Eqnref}[1]{Eq.~\eqref{eq:#1}}
\newcommand{\Eqsref}[1]{Eqs.~\eqref{eq:#1}}
\newcommand{\secref}[1]{\ref{sec:#1}}
\newcommand{\Secref}[1]{Sec.~\ref{sec:#1}}
\newcommand{\figref}[1]{\ref{fig:#1}}
\newcommand{\Figref}[1]{Fig.~\ref{fig:#1}}
\newcommand{\Figsref}[1]{Figs.~\ref{fig:#1}}

\newcommand*{\horzbar}{\rule[.5ex]{2.5ex}{0.5pt}}

\begin{document}
\title{Small-amplitude synchronisation in driven Potts models}
\author{Jan Meibohm}
\affiliation{Technische Universit\"at Berlin, Stra\ss{}e des 17. Juni 135, 10623 Berlin, Germany}
\affiliation{Department of Mathematics, King's College London, London WC2R 2LS, United Kingdom}
\author{Massimiliano Esposito}
\affiliation{Complex Systems and Statistical Mechanics, Department of Physics and Materials Science, University of Luxembourg, L-1511 Luxembourg, Luxembourg}

\begin{abstract}
We study driven $q$-state Potts models with thermodynamically consistent dynamics and global coupling. For a wide range of parameters, these models exhibit a dynamical phase transition from decoherent oscillations into a synchronised phase. Starting from a general microscopic dynamics for individual oscillators, we derive the normal form of the high-dimensional Hopf-Bifurcation that underlies the phase transition. The normal-form equations are exact in the thermodynamic limit and close to the bifurcation. Exploiting the symmetry of the model, we solve these equations and thus uncover the intricate stable synchronisation patterns of driven Potts models, characterised by a rich phase diagram. Making use of thermodynamic consistency, we show that synchronisation reduces dissipation in such a way that the most stable synchronised states dissipate the least entropy. Close to the phase transition, our findings condense into a linear dissipation-stability relation that connects entropy production with phase-space contraction, a stability measure. At finite system size, our findings suggest a minimum-dissipation principle for driven Potts models that holds arbitrarily far from equilibrium.
\end{abstract}

\maketitle

\section{Introduction}\seclab{intro}
Synchronisation is a striking collective phenomenon that occurs in systems of coupled oscillators~\cite{Pik01,Boc02,Ace05,Are08}. Due to interactions, a system of oscillators with different oscillation frequencies may phase-lock to a common frequency and oscillate coherently. Synchronisation has been observed across a range of disciplines in science~\cite{Ros03} such as biology~\cite{May72,Pes75,Mic87,Jew98}, neuroscience~\cite{Fel11}, chemistry~\cite{Tay09}, and physics~\cite{Bel19,Cal22}.

In large systems of coupled oscillators, the emergence of synchrony is an instance of a non-equilibrium phase transition, characterised by the spontaneous occurrence of spatiotemporal order -- in the form of phase locking -- through the breaking of time-translational symmetry. To this day, non-equilibrium phase transitions lack a comprehensive theoretical understanding comparable to that achieved for equilibrium phase transitions in the past century.

In the study of equilibrium phase transitions, exactly solvable models have played an important role~\cite{Bax82}. A crucial archetype of model whose equilibrium properties are mathematically accessible, is the $q$-state Potts model~\cite{Pot52,Wu82}. The model consists of $N$ identical units of $q$ different states, labelled $0,\ldots,q-1$, immersed in a heat bath at inverse temperature $\beta=1/(\kb T)$. One can think of each unit as a two dimensional vector (or spin) that points in one of $q$ equally spaced directions, see \Figref{setup}(a). In the version of the model we consider here, Potts spins interact via a global potential: The energy is reduced by $\ms{J}/N$ whenever a pair of spins is in the same state, see \Figref{setup}(b).

\begin{figure}
	\includegraphics[width=\linewidth]{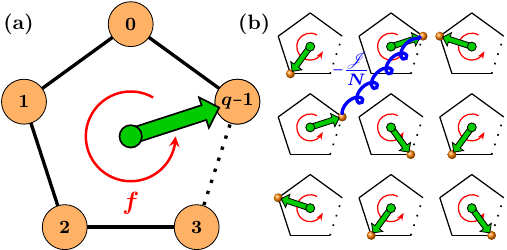}
	\caption{Sketch of the model. (a) Single Potts oscillator consisting of a two-dimensional spin (green arrow) pointing in one out of $q$ directions (orange circles) with labels $0,\ldots,q-1$. The spin jumps stochastically between adjacent states and is driven by a non-conservative force $f$ (red arrow). (b) The spins are globally coupled so that the energy $E$ is reduced by $\ms{J}/N$ for any pair of spins pointing in the same direction.}\figlab{setup}
\end{figure}

In non-equilibrium physics, exactly solvable models have played a crucial role, too.  For synchronisation, in particular, the perhaps most influential solvable model is the Kuramoto model~\cite{Kur75,Str00,Ace05}, which consists of $N\to\infty$ globally coupled oscillators with continuous phases. The natural frequency of each oscillator is fixed but random, drawn from a given probability distribution. In the Kuramoto model, the competition between decoherence, introduced by the spread in the natural frequencies, and order, introduced by an aligning interaction potential, leads to a dynamical phase transition into a synchronised state at a critical coupling strength~\cite{Kur75,Str00,Ace05}.

In this paper, we analyse driven $q$-state Potts models~\cite{Woo06,Her18,Her19}, non-equilibrium versions of the classic Potts model. In driven Potts models, the spins are amended with thermodynamically-consistent stochastic dynamics that models the interaction between the spins and the heat bath. In addition, each Potts spin is driven by a non-conservative force $f$. To model this force, the dynamics is chosen to favor one direction of rotation over the other. With such a driving, the Potts spins become ``Potts oscillators'' that, on average, rotate in the direction of the driving, see red arrows in \Figref{setup}.

\begin{figure*}
	\includegraphics[width=\linewidth]{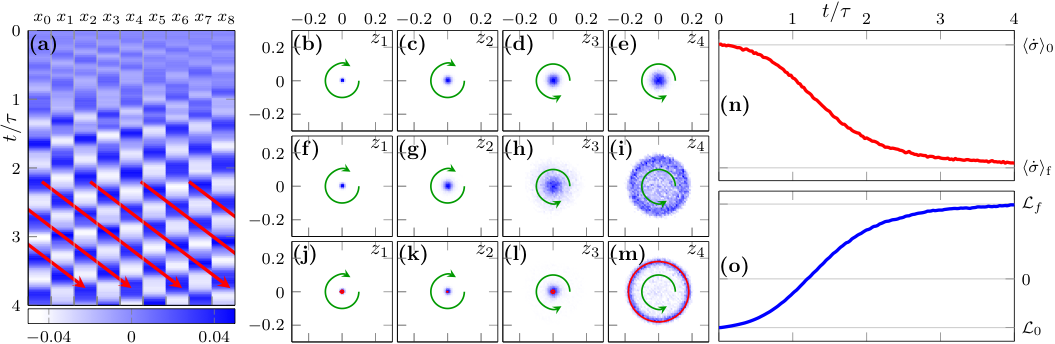}
	\caption{Oscillations in real space and Fourier space as function of time starting from the decoherent state $\ve x(0)=0$ for Arrhenius dynamics~\eqnref{Karr} with $q=9$, $\beta\!\ms{J}/q = 1.025$, $\beta f = 7$, and $N=10^5$. Arrows indicate evolution in time. (a) Deviations $\ve x(t) = (x_0,\ldots,x_8)^{\sf T}$ from decoherence as function of time, given in units of the microscopic transition time $\tau$. (b)--(m) Probability density of Fourier modes $z_1,\ldots,z_4$ obtained from an ensemble of $10^4$ realisations at $t=0.5\tau$ [(b)--(e)], $t=2\tau$ [(f)--(i)], and $t=4\tau$ [(j)--(m)]. Red lines from macroscopic dynamics. (n) Rate of dissipated work as function of time. Grey lines indicate dissipation rate in different states. (o) Phase-space contraction rate as function of time. Grey lines indicate phase-space contraction rate in different states.}\figlab{oscillations}
\end{figure*}

The explicit choice of stochastic dynamics is part of any driven Potts model and it impacts its behaviour. More generally, the dependence on dynamics is characteristic for non-equilibrium physics. At equilibrium, by contrast, the Gibbs state is universal, whenever thermodynamic consistency, i.e., detailed balance, is preserved. Away from equilibrium, no such universality exists, and even thermodynamically consistent models may be sensitive to dynamics.

For a wide range of dynamics, we show that a large number $N\to\infty$ of globally coupled Potts oscillators undergo a dynamical phase transition into a synchronised phase, in analogy with the Kuramoto model. The mechanism, however, is different here: Driven Potts models are intrinsically stochastic because they exchange energy with the heat bath. The synchronisation transition occurs due to a competition between decoherence from \textit{thermal noise} and order due to the aligning interaction between oscillators.

As we will explain in detail, large systems of Potts oscillators are described by the occupation probability $\ve p(t) = (p_0,\ldots,p_{q-1})^{\sf T}$ of the states $0,\ldots,q-1$ at time $t$. Decoherent oscillations correspond to uniform occupation, $\ve p(t) = \ve p^* \equiv (\frac1q,\ldots,\frac1q)^{\sf T}$. Figure~\figref{oscillations}(a) shows the result of numerical simulations of a large system of initially decoherent, driven Potts oscillators with given dynamics. Shown are the deviations $\ve x(t) = \ve p(t) -\ve p^*$ from decoherence for $q=9$. The initially decoherent system [$\ve x(t = 0)=0$] develops a synchronous pattern characterised by four travelling maxima, highlighted by red arrows.

Potts oscillators have an intrinsic discrete rotational symmetry, so the periodic patterns observed in \Figref{oscillations}(a) are described by discrete complex Fourier modes. As we explain below, there are four dynamical Fourier modes for $q=9$, with wave numbers $k=1,\ldots,4$, that we denote by $z_1,\ldots,z_4$. The evolution of these modes in the complex plane is shown in \Figsref{oscillations}(b)-(m).

Initially all modes are inactive [\Figsref{oscillations}(b)--(e)], indicating the initial decoherent state. As time advances, the amplitudes of the Fourier modes with largest $k$ grow [\Figsref{oscillations}(f)--(i)], until eventually [\Figsref{oscillations}(h)--(j)] only $z_4$ is active and rotates counter-clockwise in the complex plane (green arrow), reflecting the emergence of the synchronisation pattern shown in \Figref{oscillations}(a). The amplitudes of the remaining Fourier modes are essentially zero, and they rotate in different directions (green arrows). 

Thermodynamic consistency of the model enables us to extract faithful thermodynamic observables. The average rate of dissipated work, shown in \Figref{oscillations}(n), is initially at the large value $\langle \dot \sigma \rangle_0$, associated with the average dissipation rate of a single, uncoupled oscillator. During relaxation into the synchronised state, the rate of dissipated work decreases as function of time and relaxes to a smaller value $\langle\dot \sigma\rangle_\text{f}$, associated with the average dissipation rate per oscillator in the synchronised phase.

The phase-space contraction rate, shown in \Figref{oscillations}(o), is a measure of the momentary stability of a given state. Its initial value $\mathcal{L}_0$ is negative, reflecting that decoherent oscillations are unstable. But it increases as function of time, changes sign, and saturates at a positive value $\mathcal{L}_\text{f}>0$, indicating a stable synchronised final state. Comparing \Figsref{oscillations}(k) and (l) we observe an apparent connection between stability and dissipation in the system.

We explain and extend these observations by analysing analytically the equations of motion that describe the macroscopic dynamics of large systems of coupled Potts oscillators in the thermodynamic limit. So far, no analytical treatment of synchronisation in driven Potts models has been reported. The existing results are numerical, restricted to small values of $q$, and to specific choices of the stochastic dynamics~\cite{Her18,Her19}. Here we solve the model exactly close to the synchronisation transition for arbitrary dynamics and arbitrary $q$. Apart from explaining the numerical simulations shown in \Figref{oscillations}, our results reveal how the behaviour of the model depends on the choice of the dynamics and on the value of $q$.

From a linear stability analysis, we find that different dynamics may either stabilise or destabilise the decoherent phase compared to the equilibrium case. When decoherence is unstable, the model transitions into a synchronised state with coherent small-amplitude oscillations as shown in \Figref{oscillations}. We show analytically how different dynamics give rise to a variety of stable synchronisation patterns that constitute a rich phase diagram of possible synchronised states. All stable oscillations have in common that only a small number of Fourier modes is active, as observed for $q=9$ in \Figref{oscillations}(j)--(m), where only a single Fourier mode, here $z_4$, is active.

This reveals a profound difference in the way in which driven Potts models and the Kuramoto model synchronise: While ``Kuramoto-like'' synchronisation occurs in real space, ``Potts-like'' synchronisation occurs in Fourier space.

The stability of the synchronised states is closely linked to the thermodynamics of the model, as we discuss in detail in Ref.~\cite{Mei24a} that appears together with this paper. We find that synchronisation in driven Potts models reduces dissipation in such a way that the most stable synchronised states dissipate the least entropy. Close to the synchronisation transition, we derive a linear stability-dissipation relation between states of least entropy production and states of largest phase-space contraction. As we argue in detail, this finding suggests a minimum-dissipation principle~\cite{Pri55} for driven Potts models at finite $N$, that holds arbitrarily far from equilibrium and independently of the stochastic dynamics.

Equilibrium Potts models have been realised experimentally for $q=3,4$ using, e.g., spins in adsorbed systems~\cite{Bar78,Roe81,Sch81} and fluid mixtures~\cite{Das79}. Synchronisation of oscillating spins was achieved by applying an external torque through a spin-polarised current~\cite{Tso00,Man05,Awa17}. Idealised models have proven instrumental in predicting the outcomes of these experiments~\cite{Muk76,Dom77,Slo96,Ber96}. Combinations of these existing experimental approaches thus offer interesting future perspectives for the experimental observation of the Potts-like synchronisation established here.

This paper is organised as follows: In \Secref{model} we explain driven Potts models in detail, starting from the dynamics of a single Potts oscillator. We discuss the observables that characterise the thermodynamics and the stability of the model for a finite number $N$ of oscillators. In \Secref{thermo} we introduce intensive parameters and take the thermodynamic limit $N\to\infty$. In this limit, we derive the deterministic equations of motion for the macroscopic dynamics of the occupation probability $\ve p(t)$ of oscillators. We discuss the stability of the decoherent phase in \Secref{stabdec}. In \Secref{saoscil} we analyse of the stability and thermodynamics of synchronised and stationary states close to the phase transition. To this end, we derive the normal form of the Hopf bifurcation, for which analysis results in the phase diagram of the model. We then connect the stability of these states to their dissipation. In \Secref{numsim} we present numerical simulations for large but finite $N$ and compare them to the theory. Conclusions are drawn in \Secref{conc}. Appendices~\secref{appctrafo}--\secref{apppscontr} contain detailed calculations to support the main text.
\section{Model}\seclab{model}
In this first Section, we give a detailed description of driven Potts models. We start by introducing a single uncoupled Potts oscillator, followed by a discussion of $N$ uncoupled oscillators. In the last step, we consider the complete model of $N$ interacting Potts oscillators with thermodynamically consistent stochastic dynamics.
\subsection{Single uncoupled oscillator}
When the interaction strength $\ms{J}$ is set to zero, one may consider each Potts oscillator individually. Each oscillator $s_{m}$ with index $m=1,\ldots,N$ can occupy one of $q$ states
\algn{
	s_{m} \in \{0,\ldots,q-1\}\,.
}
Microscopic transitions within an oscillator are induced by thermal fluctuations of the surrounding heat bath. Transitions are allowed only between adjacent states, i.e. only for 
\algn{\eqnlab{alltrans}
	s_{m}\to s_{m}\pm1\,,
}
where $s_{m}+1$ corresponds to a counter-clockwise transition and $s_{m}-1$ to a clockwise transition. All states are understood modulo $q$ to implement the periodicity of Potts spins.

Fluctuations of the heat bath are assumed to occur on much shorter time scales than the dynamics of the system. We therefore model transitions~\eqnref{alltrans} by a Markovian stochastic dynamics with transition rates $k^\pm$, that obey local detailed balance~\cite{Sei12,Bro15,Pel21}
\algn{\eqnlab{ldbsingle}
 	\frac{k^\pm}{k^\mp} = \ee^{\pm\beta f}\,.
}
Here, $f$ and $\beta$ denote the non-conservative force and the inverse temperature of the heat bath, respectively, introduced in \Secref{intro}. Positive driving $f>0$ favours counter-clockwise over clockwise transitions, leading to a mean probability flux in counter-clockwise direction, the direction of the driving. Local detailed balance~\eqnref{ldbsingle} ensures that the model is amenable to a consistent thermodynamic description in the framework of Stochastic Thermodynamics~\cite{Sei12,Bro15,Pel21}.

The probability $p_n$ of finding a single, uncoupled oscillator $s_{m}$ in state $n$ obeys the Master equation~\cite{Kam07,Gar09}
\algn{\eqnlab{meq}
	\dot p_n = %k^+p_{n-1} + k^-p_{n+1} - (k^+ + k^-)p_n\,.
	j(p_n,p_{n-1})\big|_{\ms{J}=0} - j(p_{n+1},p_n)\big|_{\ms{J}=0}\,,
}
where $j(p_{n+1},p_n)\big|_{\ms{J}=0}$ is the probability flux from state $n$ to state $n+1$ at vanishing coupling:
\algn{\eqnlab{singleflux}
	j(p_{n+1},p_n)\big|_{\ms{J}=0} = k^+p_n - k^-p_{n+1}\,.
}
Due to discrete rotational symmetry, \Eqnref{meq} is diagonalised by discrete Fourier transform. The Fourier modes
\algn{
	\hat p_k = \sum_{n=0}^{q-1}\ee^{\frac{i2\pi k n}{q}}p_n\,,
}
obey the transformed Master equation
\algn{
	\dot{\hat p}_k = \left(\mu_k|_{\ms{J}=0} + i\omega_k\big|_{\ms{J}=0}\right)\hat{p}_k\,,
}
where $\mu_k|_{\ms{J}=0}\leq0$ represents the damping at vanishing coupling and $\omega_k$ is the angular frequency of the $k^\text{th}$ Fourier mode,
\sbeqs{\eqnlab{muomsingle}
\algn{
	\mu_k|_{\ms{J}=0} =& -2(k^++k^-)\sin^2\left(\frac{\pi k}{q}\right)\,,\\
	\omega_k|_{\ms{J}=0} =& (k^+-k^-)\sin\left(\frac{2\pi k}{q}\right)\,.
}
}
The coefficients $\mu_k|_{\ms{J}=0}$ are negative for all $k$ except $k=0$, as a consequence of the Perron-Frobenius theorem~\cite{Per07,Fro12}. Vanishing $k=0$ corresponds to the decoherent state
\algn{\eqnlab{decstate}
	\ve p^*\equiv\left(\frac1q,\dots,\frac1q\right)^{\sf T}\,,
}
introduced in \Secref{intro}, for which the probabilities are homogeneously distributed among all states $0,\ldots,q-1$.  This is the unique steady state for a single, uncoupled Potts oscillator.

As we discuss in \Secref{stabdec} below, interactions between oscillators may change the sign of $\mu_k$ when $\beta\!\!\ms{J}$ reaches a critical value, rendering the decoherent phase unstable. The angular frequencies $\omega_k$, by contrast, are only weakly affected by interactions once the decoherent phase is unstable. In analogy with the natural frequencies of individual oscillators in the Kuramoto model, we call $\omega_k$ the ``natural frequencies'' of driven Potts models.
\subsection{$N$ uncoupled oscillators}\seclab{uncoup}
In the next step, we consider the behaviour of an ensemble of $N$ uncoupled, identical oscillators. This system is invariant under arbitrary permutations of oscillators. It is therefore convenient to introduce the $q$-dimensional occupation vector $\ve N = (N_0,\ldots,N_{q-1})^{\sf T}$ associated with a given microstate $\ve s$. The components $N_n$ of $\ve N$ specify the number of oscillators in state $n=0,1,\ldots,q-1$ and are subject to the constraint $\sum_{n=0}^{q-1} N_n=N$.

The occupation vector $\ve N$ provides a coarse-grained but complete description of the microstate. However, since the coarse-graining map $\ve s \mapsto \ve N$ is many-to-one, the states described by $\ve N$ carry a (dimensionless) internal entropy~\cite{Her18,Her19,Her20}
\algn{\eqnlab{internalentropy}
	S_\text{int}(\ve N)=\log\Omega(\ve N)\quad \text{with} \quad\Omega(\ve N) =\frac{N!}{\prod_{n=0}^{q-1} N_n!}\,,
}
where $\Omega$ is the microscopic multiplicity of the state $\ve N$. Here and in the following we express all entropy measures in units of $\kb$.

An allowed microscopic transition~\eqnref{alltrans} for $s_m=n$ leads to a change of the occupation vector
\algn{\eqnlab{alltransN}
	\ve N\to \ve N+\ve \Delta^\pm_n\quad \text{with}\quad (\ve \Delta^\pm_n)_m = \delta_{mn\pm1}-\delta_{mn}\,,
}
where $\delta_{mn}$ denotes the Kronecker symbol. The inverse transitions of $\ve \Delta^\pm_n$ are given by
\algn{
	\ve \Delta^\mp_{n\pm1} = -\ve \Delta^\pm_n\,.
}
In the non-interacting model, microscopic transitions occur independently. Therefore, the rates $W^\pm_n(\ve N)\big|_{\ms{J}=0}$ for the transitions~\eqnref{alltransN} are the sums of the rates of all microscopic transitions that enter~\eqnref{alltransN}, i.e.,
\algn{\eqnlab{transratesnonint}
	W^\pm_n(\ve N)\big|_{\ms{J}=0} =N_n k^\pm\,.
}
Consequently, the transition rates \eqnref{transratesnonint} of the non-interacting model are products of the entropic factors $N_n$, that stem from the combined contributions of individual microscopic oscillators in state $n$, and the microscopic transition rates $k^\pm$.

The transition rates~\eqnref{transratesnonint} on the level of the occupation vector $\ve N$, obey a macroscopic version of the microscopic local detailed balance condition~\eqnref{ldbsingle}:
\algn{\eqnlab{ldbnonint}
	\frac{W^\pm_n(\ve N)}{W^\mp_{n\pm1}(\ve N+{\ve \Delta}_{n}^\pm)}\bigg|_{\ms{J}=0}\hspace{-3.5mm} = \exp\left[S_\text{int}(\ve N+\ve \Delta^\pm_n)-S_\text{int}(\ve N)\pm \beta f	\right],
}
that involves the change in internal entropy
\algn{
	S_\text{int}(\ve N+\ve \Delta^\pm_n)-S_\text{int}(\ve N) = \log\left(\frac{N_n}{N_{n\pm1}+1}\right)\,,
}
along the transition.

Given the transition rates~\eqnref{transratesnonint}, the probability $P(\ve N,t)$ to find the system in state $\ve N$ at time $t$ obeys the Master equation
\algn{\eqnlab{mastereqnnonint}
	\dot P(\ve N,t) 	&= \sum_{n=0}^{q-1}\left[J_n(\ve N, \ve N -{\ve \Delta}_n^+)\big|_{\ms{J}=0}-J_n(\ve N+{\ve \Delta}_n^+, \ve N)\big|_{\ms{J}=0}\right]\,,
}
where
\algn{
	J_n(\ve N',\ve N)\big|_{\ms{J}=0} = N'_n k^+P(\ve N',t) - N_{n+1}k^-P(\ve N,t)\,,
}
denotes the probability flux from $\ve N$ to $\ve N'$ along state $n$.
\subsection{Complete model}
In the final step, we now discuss the complete driven Potts model, including interactions. The oscillators interact with each other pairwise by a global, ferromagnetic interaction of strength $\ms{J}/N$, whenever two oscillators are in the same state, see \Figref{setup}(b). In other words, every microscopic pair $(m,m')$ with $s_m = s_{m'}$, $m\neq m'$, reduces the internal energy $E$ of the system by $\ms{J}/N$.

The global nature of the interaction conserves the invariance of the system under arbitrary permutations of the oscillators, so we can describe the complete model using the occupation vector $\ve N$ introduced in \Secref{uncoup}. A simple calculation shows that the total energy $E$ is expressed in terms of $\ve N$ as
\algn{\eqnlab{epot}
 	E(\ve N) = - \frac{\ms{J}}{2N}\left(\ve N\cdot \ve N - N\right)\,.
}

Analogous to the non-interacting rates~\eqnref{transratesnonint}, the transition rates $W^\pm_n(\ve N)$ for the transitions $\ve N\to \ve N + \ve \Delta^\pm_n$ of the interacting model,
\algn{\eqnlab{transrates}
	W^\pm_n(\ve N) =N_n K^\pm(N_{n\pm1}-N_n)\,,
}
are products of $N_n$ and the functions $K^\pm(x)$ that contain the microscopic interactions between spins as well as the non-conservative force. We assume that the functions $K^\pm(x)$ depend on the energy change due to the transition,
\algn{\eqnlab{energychange}
	E(\ve N + \ve \Delta^\pm_n)-E(\ve N) & = -\frac{\ms{J}}N\left(N_{n\pm1}-N_n +1	\right)\,,
}
but that they are otherwise independent of $n$. In contrast to the extensive total energy~\eqnref{epot}, the energy change~\eqnref{energychange} is intensive in the thermodynamic limit $N\to\infty$. Therefore, the dependence of $K^\pm(x)$ on the energy change~\eqnref{energychange} ensures a consistent thermodynamic limit for the transition rates. The argument of $K^\pm(x)$, i.e., $x=N_{n\pm1}-N_n$ in \Eqnref{transrates}, then follows from the $\ve N$ dependence given in \Eqnref{energychange}.

Since the occupation vector $\ve N$ carries an internal entropy $S_\text{int}(\ve N)$, the free energy $F$ of the system~\cite{Her18,Her19,Her20}
\algn{\eqnlab{freeen}
	F(\ve N) = E(\ve N) - \beta^{-1}S_\text{int}(\ve N)\,,
}
enters the local detailed balance condition for the transition rates of the interacting model, i.e.,
\algn{\eqnlab{ldb}
	\frac{W^\pm_n(\ve N)}{W^\mp_{n\pm1}(\ve N+\ve \Delta^\pm_n)} =\exp\left\{-\beta\left[ F(\ve N+\ve \Delta^\pm_n) - F(\ve N)\mp f	\right]\right\}\,.
}
The product form~\eqnref{transrates} of the transition rates further implies a local detailed balance condition for $K^\pm(x)$ with respect to the energy change alone,
\algn{\eqnlab{ldben}
	\frac{K^\pm(N_{n\pm1}-N_n)}{K^\mp(N_n-N_{n\pm1})} =\exp\left\{-\beta\left[ E(\ve N+\ve \Delta^\pm_n) - E(\ve N)\mp f	\right]\right\}\,.
}

In the non-interacting limit $\ms{J}\to0$, the rates $K^\pm(x)$ reduce to the $\ve N$-independent $k^\pm$. Consequently, \Eqnref{ldb} reduces to \eqnref{ldbnonint} and \Eqnref{ldben} reduces to \eqnref{ldbsingle}.

Finally, the Master equation for the probability $P(\ve N,t)$ of the complete model reads
\algn{\eqnlab{mastereqn}
	\dot P(\ve N,t) 	&= \sum_{n=0}^{q-1}\left[J_n(\ve N, \ve N -{\ve \Delta}_n^+)-J_n(\ve N+{\ve \Delta}_n^+, \ve N)\right]\,,
}
with probability fluxes for the transitions $\ve N\to\ve N'$ along $n$ given by
\algn{
	J_n(\ve N',\ve N) = W^+_n(\ve N)P(\ve N,t) - W^-_{n+1}(\ve N')P(\ve N',t)\,.
}
\subsection{Choice of dynamics}
The local detailed balance conditions~\eqnref{ldb} and \eqnref{ldben} fix the antisymmetric part of the logarithms of $K^\pm(x)$, but leave them otherwise arbitrary. Ideally, one would have a microscopic model for the (e.g. quantum mechanical) interaction between the heat bath and the Potts spins, that would allow one to derive $K^{\pm}(x)$ explicitly. However, such a model is not known in general, which makes the freedom in choosing $K^{\pm}(x)$ an integral part of Potts models. The best one can do is to regard the undetermined (symmetric) part of the logarithm of $K^{\pm}(x)$ as a free parameter of the model and to explore the parameter space that this choice offers.

While it is possible to make general statements about the non-equilibrium properties of driven Potts models close to the synchronisation transition, as we will show, numerical simulations require specific choices of the transition rates. A typical choice uses mixed Arrhenius rates with
\algn{\eqnlab{Karr}
	K_\text{Arr}^{\pm}(x) = \tau^{-1}\ee^{\frac{\beta}{2}\left[\frac{\ms{J}}N(1\mp\xi)\left(x +1	\right)\pm f\right]}\,,
}
depending on the parameter $-1\leq \xi \leq 1$~\footnote{We note that the used term ``Arrhenius rate'' refers to the exponential form of the transition rates in \Eqnref{Karr} and not to its dependence on the initial and final energies.}. The time scale $\tau$ corresponds to the typical, microscopic transition time of a single Potts spin. For $\xi=0$, $K_\text{Arr}^{\pm}(x)$ correspond to the classic Arrhenius rates, studied in Refs.~\cite{Her18,Her19}, and used in the numerical simulations presented in \Figref{oscillations}.

A second common choice uses Glauber rates
\algn{\eqnlab{Kgla}
	K_\text{Gla}^{\pm}(x) = \frac2{\tau}\frac{\ee^{\pm \frac{\beta f}2 }}{\ee^{-\frac{\ms{J}}N\left(x +1	\right)}+1}\,.
}
Arrhenius rates~\eqnref{Karr} with $\xi=0$ and Glauber rates~\eqnref{Kgla} have convenient properties, that we discuss in more detail in \Secref{stabexp}. For the numerical simulations presented in \Secref{numsim}, we therefore use either of these two. Notwithstanding this, our analytical results hold in general and thus for all dynamics.
\subsection{Thermodynamic observables}\seclab{obs}
Thermodynamic consistency of the model in the form of the local detailed balance conditions~\eqnref{ldb} and~\eqnref{ldben}, constrains the dynamics and allows us to define genuine stochastic thermodynamic quantities at the mesoscopic level, that are consistent with the macroscopic thermodynamics of the system~\cite{Fal23}.

In particular, the system entropy $S_\text{sys}$ is given by the sum of the internal entropy and Shannon entropy~\footnote{The additive form of the $S_\text{sys}$ is a direct consequence of the coarse graining $\ve s\mapsto\ve N$, see e.g. the discussion in Ref.~\cite{Sei19}.}:
\algn{\eqnlab{ssys}
	S_\text{sys}(\ve N,t)  =  S_\text{int}(\ve N) + S_\text{sha}(\ve N,t)\,,
}
where the stochastic Shannon entropy reads
\algn{
	S_\text{sha}(\ve N,t) = -\log P(\ve N,t)\,.
}
Averaging $S_\text{sys}$ over $\ve N$ and taking a time derivative, we write the average rate of change of system entropy $\langle \dot S_\text{sys} \rangle$ in terms of the average total entropy production rate $\langle\dot\Sigma\rangle$ and the rate of entropy change $\langle\dot S_\text{env}\rangle$ of the environment:
\algn{
	\langle \dot S_\text{sys} \rangle = \langle \dot\Sigma \rangle - \langle \dot S_\text{env}\rangle\,,
}
where $\langle\cdot\rangle$ denotes the average over $\ve N$. These quantities are expressed in terms of probabilities, fluxes, and rates as
\sbeqs{\eqnlab{stochobs}
\algn{
	&\langle \dot\Sigma \rangle = \sum_{\ve N}\sum_{n=0}^{q-1}\log\left[\frac{W_n^+(\ve N)P(\ve N,t)}{W_{n+1}^{-}(\ve N+\ve \Delta_n^+)P(\ve N+\ve \Delta_n^+,t)}\right]J_n(\ve N+{\ve \Delta}_n^+, \ve N),\\
	&\langle\dot S_\text{env} \rangle = \sum_{\ve N}\sum_{n=0}^{q-1}\log\left[\frac{K^+(N_{n+1}-N_n)}{K^-(N_n-N_{n+1})} \right]J_n(\ve N+{\ve \Delta}_n^+, \ve N)\,.
}

With the local detailed balance condition~\eqnref{ldben}, we rewrite $\langle\dot S_\text{env}\rangle$ in terms of the energy change and the non-conservative force as
\algn{
	\langle \dot S_\text{env} \rangle &= -\beta\sum_{\ve N}\sum_{n=0}^{q-1}\left[ E(\ve N+\ve \Delta^+_n) - E(\ve N)-f	\right]J_n(\ve N+{\ve \Delta}_n^+, \ve N)\,,\nn\\
							&= -\beta\langle \dot Q\rangle\,,\eqnlab{SenvQ}
}
}
where $\langle \dot Q\rangle$ denotes the average heat flux into the system.

The first law of thermodynamics, $\langle \dot E\rangle = \langle \dot Q\rangle + \langle \dot W\rangle$, relates the average heat flux $\langle \dot Q\rangle$ to the average rate of energy change $\langle \dot E\rangle$ and the average rate of dissipated work $\langle \dot W\rangle$, identified as
\algn{
	\langle \dot W\rangle = f\sum_{\ve N}\sum_{n=0}^{q-1}J_n(\ve N+{\ve \Delta}_n^+, \ve N)\,.
}

All three quantities, $\langle \dot\Sigma \rangle$, $\langle\dot S_\text{env} \rangle$, and $\langle \dot W\rangle$, are measures of dissipation. The total entropy production rate $\langle \dot\Sigma \rangle$ determines the irreversibility of the complete system of oscillators and heat bath as function of time and is therefore arguably the most physically relevant dissipation measure. On the other hand, the other two measures, the average rate of dissipated work $\langle \dot W\rangle$ in particular, are much easier to compute along stochastic trajectories, because their expressions [cf. \Eqsref{stochobs}] do not explicitly involve the probability $P(\ve N,t)$.

Fortunately, at steady state, i.e. $\dot P(\ve N,t) = 0$, we have $\langle \dot E \rangle = \langle \dot S_{\text{sys}}\rangle = 0$, which implies that
\algn{\eqnlab{ssrel}
	\langle \dot \Sigma \rangle =  \langle \dot S_\text{env} \rangle = \beta\langle \dot W\rangle.
}
This shows that the average rate of dissipated work $\langle \dot W\rangle$ can be used as a proxy for the average entropy production rate $\langle\dot\Sigma\rangle$ at steady state. In the thermodynamic limit, discussed in \Secref{thermo}, an equation similar to~\eqnref{ssrel} holds even for non-steady states at long times, so the average rate of dissipated work is a faithful measure of dissipation in this case as well.
\subsection{Inflow rate}
In addition to dissipation, measured by the thermodynamic observables introduced in \Secref{obs}, we now consider the stability of trajectories. A simple measure for the stability of a stochastic trajectory is the so-called inflow rate~\cite{Bai15}, defined as the entrance rate $W_\text{in}(\ve N)$, the rate of trajectories entering a given state $\ve N$, subtracted by the escape rate $W_\text{out}(\ve N)$ out of $\ve N$:
\algn{\eqnlab{entrate}
	L(\ve N) = W_\text{in}(\ve N) - W_\text{out}(\ve N)\,,
}
where
\sbeqs{
\algn{
	W_\text{in}(\ve N) &= \sum_{n=0}^{q-1}\left[W^+_n(\ve N - \ve{\Delta}_n^+)+W^-_n(\ve N - \ve{\Delta}_n^-)\right]\,,\\
	W_\text{out}(\ve N) &= \sum_{n=0}^{q-1}\left[W^+_n(\ve N)+W^-_n(\ve N)\right]\,.
}
}
The inflow rate quantifies the tendency of probability to accumulate in $\ve N$ when $L(\ve N)>0$, and to be depleted when $L(\ve N)<0$. As we discuss in \Secref{thermo}, in the thermodynamic limit $N\to\infty$ the average $\langle L\rangle$ of the (stochastic) inflow rate $L(\ve N)$ reduces to the (deterministic) phase-space contraction rate of the macroscopic dynamics, a measure of stability in dynamical systems theory~\cite{Ott02}.
\section{Thermodynamic limit}\seclab{thermo}
In the thermodynamic limit $N\to\infty$, the law of large numbers ensures that intensive variables are well described by their mean values.

In particular, the rescaled occupation number $\ve n  \equiv \ve N/N$ is intensive and obeys such a law. As a consequence, fluctuations of $\ve n$ become unimportant in the thermodynamic limit and $\ve n$ converges to its (deterministic) mean value $\ve n \to \ve p(t) \equiv \langle \ve n\rangle$~\cite{Kam07}. The quantity $p_n(t)= \langle n_n\rangle$, in turn, gives the probability that a randomly chosen microscopic Potts oscillator $s_m$ is in state $s_m=n$ at time $t$.

Because fluctuations of $\ve n$ vanish as $N\to\infty$, intensive averages of functions of $\ve n$ have large-$N$ limits that are deterministic functions of $\ve p$. For a given extensive quantity $A(\ve N)$ with the intensive limit $a(\ve p)$, we have
\algn{\eqnlab{mfaverage}
	\lim_{N\to\infty}\frac1N\langle A(\ve N)\rangle =  \lim_{N\to\infty}\frac1N A(\langle\ve n\rangle N) \equiv a(\ve p)\,.
}
Similarly, averages of ratios of two extensive quantities $A(\ve N)$ and $B(\ve N)$, with individual intensive limits $a(\ve p)$ and $b(\ve p)$, decompose as
\algn{\eqnlab{mfratios}
	\lim_{N\to\infty}\Big\langle\frac{A(\ve N)}{B(\ve N)}\Big\rangle =  \lim_{N\to\infty}\frac{A(\langle\ve n\rangle N)}{N}\left(\frac{B(\langle\ve n\rangle N)}{N}\right)^{-1} =\frac{a(\ve p)}{b(\ve p)}\,.
}
Using \Eqnref{mfaverage}, we find the large-$N$ limits
\sbeqs{
\algn{
	\lim_{N\to\infty}\frac1N\langle E(\ve n N)\rangle&= \lim_{N\to\infty}\frac1N E(\langle\ve n\rangle N)  \equiv \mc{E}(\ve p)\,,\\
	 \lim_{N\to\infty}\frac1{N}\langle S_\text{int}(\ve n N)\rangle &= \lim_{N\to\infty}\frac1{N} S_\text{int}(\langle\ve n\rangle N)\equiv\mc{S}_\text{int}(\ve p)\,,\\
	\lim_{N\to\infty}\frac{1}{N}\langle F(\ve n N)\rangle &= \lim_{N\to\infty}\frac{1}{N} F(\langle\ve n\rangle N) \equiv\mc{F}(\ve p)\,, 
}
}
where
\algn{
 	\mc{E}(\ve p) 			&= -\frac{\ms{J}}{2}\ve p\cdot\ve p\,,\qquad \mc{S}_\text{int}(\ve p)	= -\ve p\cdot\log \ve p\,,\\
	\mc{F}(\ve p)			&= \mc{E}(\ve p) - \beta^{-1}\mc{S}(\ve p)\,. \eqnlab{freeendens}
}
Here, $\mc{E}(\ve p)$ denotes the mean energy per oscillator, $\mc{S}_\text{int}(\ve p)$ is the mean internal entropy per oscillator, and $\mc{F}(\ve p)$ represents the mean free energy per oscillator.

The $N$-rescaled, average fluxes are also intensive and read
\algn{\eqnlab{avflux}
	\frac{1}N\sum_{\ve N}J_n(\ve N+\ve{\Delta}^+_n,\ve N) \equiv \frac1N\left[\langle W^+_n(\ve n N)\rangle - \langle W^-_{n+1}(\ve n N)\rangle\right]\,.
}
As $N\to\infty$ the terms on the right-hand side of \Eqnref{avflux} become [cf. \Eqnref{mfaverage}]
\algn{
	\lim_{N\to\infty}\frac1N\langle W^{\pm}_n(\ve n N)\rangle = \lim_{N\to\infty}\frac{W^{\pm}_n(\langle\ve n\rangle N)}{N} \equiv w_n^\pm(\ve p )\,,
}
where the rescaled transition rates $w_n^\pm$ are given by
\algn{\eqnlab{rates}
	w_n^\pm(\ve p ) = k^\pm(p_{n\pm1}-p_n)p_n \,,
}
and
\algn{
	\quad k^\pm(p_{n\pm1}-p_n) = \lim_{N\to\infty}\langle K^{\pm}[(n_{n\pm1} - n_{n})N]\rangle\,.
}
For the thermodynamic limit of the rescaled fluxes in \Eqnref{avflux}, we then obtain
\algn{
	j(p_{n+1},p_n) &\equiv \lim_{N\to\infty} \frac{1}N\sum_{\ve N}J_n(\ve N+\ve{\Delta}^+_n,\ve N)\,,\nn\\
	 &= k^+(p_{n+1}-p_n)p_n - k^-(p_{n}-p_{n+1})p_{n+1}\,. \eqnlab{flux}
}
The function $j(p_{n+1},p_n)$ represents the average probability flux per oscillator from state $n$ to state $n+1$ in the thermodynamic limit.

Note that the average flux $j(x',x)$ in \Eqnref{flux} has the same form as the flux~\eqnref{singleflux} of a single oscillator. The difference is that in the interacting model with $\ms{J}>0$, the functions $k^\pm(x)$ in \Eqnref{flux} depend on $\ve p$.
\subsection{Macroscopic dynamics}
We now consider the dynamics of $\ve p = \langle \ve n\rangle$ in the thermodynamic limit. To this end, we multiply \Eqnref{mastereqn} by $\ve n = \ve N/N$ and average over $\ve N$. Using the same arguments as before, we obtain for $N\to\infty$ the equation of motion
\algn{\eqnlab{mf}
	\dd{t}\ve p(t) 	\equiv	 \ve h[\ve p(t)] = \sum_{n=0}^{q-1} \left[\ve \Delta^+_n w_n^+(\ve p) + \ve \Delta_n^-w_n^-(\ve p )\right]\,.
}
To simplify \Eqnref{mf}, we use \Eqsref{alltransN} to remove the sum and express the right-hand side in terms of the average fluxes~\eqnref{flux}. This gives
\algn{\eqnlab{eom}
	\dot p_n = h_n(\ve p) =  j(p_{n},p_{n-1}) - j(p_{n+1},p_n)\,,
}
with all indices understood modulo $q$.

Again, \Eqnref{eom} has the same form as Master equation~\eqnref{meq} for a single, uncoupled oscillator. And just as for the linear dynamics \eqnref{meq}, one can readily show that \Eqnref{eom} conserves probability, i.e.,
\algn{\eqnlab{constr}
	\dd{t} \sum_{n=0}^{q-1} p_n(t) = 0\,,\quad \sum_{n=0}^{q-1} p_n(t) = 1\,.
}
Due to the interactions $\ms{J}>0$ between Potts oscillators, however, the fluxes $j(y,x)$ in~\eqnref{eom} are non-linear functions of the probability $\ve p$, whereas \Eqnref{meq} is linear and constrained by the Perron-Frobenius theorem~\cite{Per07,Fro12}.

The macroscopic dynamics is inherently thermodynamically consistent and satisfies a macroscopic version of local detailed balance condition~\eqnref{ldb}:
\algn{\eqnlab{ldbmf}
	\frac{w^\pm_n(\ve p)}{w^\mp_{n\pm1}(\ve p )}	&=\exp\left\{-\beta \left[\ve{\Delta}^\pm_n\cdot\ve{\nabla}_{\!\!\ve p}\mc{F}(\ve p)\mp f\right]\right\}\,,
}
which follows from applying \Eqnref{mfratios} to \Eqnref{ldb}. The change in free energy enters \Eqnref{ldbmf} as
\algn{
	\ve{\Delta}^\pm_n\cdot\ve{\nabla}_{\!\!\ve p}\mc{F}(\ve p)&=\lim_{N\to\infty}\frac1N\left\{\langle F[ (\ve n + \ve \Delta^\pm_k/N)N]\rangle-\langle F(\ve n N)\rangle\right\} \,,\nn\\
	&= -\ms{J}(p_{n\pm1}-p_n)+\beta^{-1}\left(\log p_{n\pm1} - \log p_n\right)\,.
}
Here, $(\nabla_{\ve p})_n\equiv \frac{\partial}{\partial p_n}$ denotes the gradient with respect to $\ve p$.
\subsection{Macroscopic thermodynamic observables}
The intensive versions of the thermodynamic observables introduced in \Secref{obs} have large-$N$ limits in terms of the probabilities $\ve p$.

We define the intensive entropy production rate $\langle\dot\sigma\rangle = \lim_{N\to\infty}\langle\dot\Sigma\rangle/N$, rate of entropy change of the environment $\langle\mc{\dot S}_\text{env}\rangle=\lim_{N\to\infty}\langle \dot S_\text{env}\rangle/N$, rate of dissipated work $\langle\mc{\dot W}\rangle = \lim_{N\to\infty}\langle \dot W\rangle/N$, and heat production rate $\langle\mc{\dot Q}\rangle = \lim_{N\to\infty}\langle \dot Q\rangle/N$.

From the thermodynamic limit of \Eqsref{stochobs}, we find that the intensive thermodynamic observables $\langle\mc{\dot O}\rangle=\langle\mc{\dot W}\rangle,\langle\mc{\dot S}_\text{env}\rangle,\langle\dot\sigma\rangle$ can be decomposed into
\algn{\eqnlab{compobs}
	\langle \mc{\dot O} \rangle =& \sum_{n=0}^{q-1}\mc{\dot O}(p_{n+1},p_n)\,,
}
with edge functions $\mc{\dot O}(y,x)$ that are identical for all $n$. The edge functions read in terms of the average flux $j(y,x)$:
\sbeqs{\eqnlab{obsform}
\algn{
	\dot \sigma (y,x) 	=& \left[\beta\!\!\ms{J}(y-x)+\beta f+\log\left(x/ y\right)\right]j(y,x)\,,\\
	\mc{\dot S}_\text{env}(y,x) =& \beta \left[\ms{J}(y-x)+f\right]j(y,x)\,,\\
	\mc{\dot W}(y,x) =& fj(y,x)\,.
}
}
Using these explicit forms, we write $\langle \mc{\dot S}_\text{env} \rangle$ and $\langle \dot\sigma \rangle$ in terms of $\langle \mc{\dot W}\rangle$ as
\sbeqs{\eqnlab{obstder}
\algn{
	\langle \dot \sigma \rangle =& \beta \langle \mc{\dot W} \rangle-\beta\dd{t}\mc{F}[\ve p(t)]\,,\\
	\langle \mc{\dot S}_\text{env} \rangle =& \beta\langle \mc{\dot W}\rangle-\beta\dd{t}\mc{E}[\ve p(t)]\,.
}
}
These relations allow us to generalise \Eqnref{ssrel} outside the steady state in the thermodynamic limit. To this end, we take an additional long-time average over the observables
\algn{\eqnlab{tav}
	\llangle\mc{\dot O}\rrangle = \lim_{T\to\infty}\frac1{T}\int_0^{T}\!\!\ed t\langle \mc{\dot O} \rangle\,.
}
The integral in \Eqnref{tav} turns the total derivatives in \Eqsref{obstder} into boundary terms that vanish in the long-time limit. With the additional time average, we obtain a generalisation of \Eqnref{ssrel},
\algn{\eqnlab{tavrel}
	\llangle \dot \sigma \rrangle =  \llangle \mc{\dot S}_\text{env} \rrangle = \beta\llangle \mc{\dot W}\rrangle\,,
}
that is valid beyond steady states.

In the synchronised phase, when $\ve p$ follows a periodic trajectory with period $T_\text{p}$, such that $\ve p(t) = \ve p(t+T_\text{p})$, the long-time average in \Eqnref{tav} can be replaced by an average over one period:
\algn{
	\llangle\mc{\dot O}\rrangle = \frac1{T_\text{p}}\int_0^{T_\text{p}}\!\!\ed t\langle \mc{\dot O} \rangle\,.
}
Hence, with the appropriate averages, relations~\eqnref{tavrel} and \eqnref{ssrel} imply that the average rate of dissipated work $\llangle\mc{\dot W}\rrangle$ serves as convenient proxy for the dissipation of the system.
\subsection{Phase-space contraction in thermodynamic limit}
The average $\langle L\rangle$ of the entrance rate $L(\ve N)$ in \Eqnref{entrate} possesses a well-defined thermodynamic limit
\algn{
	\mc{L} \equiv \lim_{N\to\infty}\langle L(\ve N)\rangle = \lim_{N\to\infty}\langle W_\text{in}(\ve N) - W_\text{out}(\ve N)\rangle\,,
}
that reads
\algn{
	\mc{L}(\ve p) = -{\ve \nabla}_{\!\!\ve p}\cdot\sum_{n=0}^{q-1} \left[\ve \Delta^+_n w_n^+(\ve p) + \ve \Delta_n^-w_n^-(\ve p )\right]= -{\ve \nabla}_{\!\!\ve p}\cdot\dot{\ve p}\,.
}
Hence, the average entrance rate $\mc{L}$ converges to the negative flow divergence $-{\ve \nabla}_{\!\!\ve p}\cdot\ve{h}$ of $\ve p$ as $N\to\infty$. The flow divergence determines the exponential contraction rate of infinitesimal volumes $\mc{V}$ in the phase-space of $\ve p$~\cite{Ott02}. One has
\algn{
	\dd{t}\log\mc{V}(t) = -\mc{L}(\ve p)\,,
}
which is why we call $\mc{L}(\ve p)$ the phase-space contraction rate.

Phase-space probability accumulates in regions of positive $\mc{L}$, while it escapes regions of negative $\mc{L}$. Positive $\mc{L}$ is a necessary condition for flow-invariant sets, such as fixed points and for periodic orbits to be stable, and thus to be approached in the long-time limit. These properties make $\mc{L}$ (and $\langle L\rangle$) important measures of stability. 
\subsection{Symmetries of macroscopic dynamics}\seclab{symm}
In addition to the global permutation symmetry among oscillators that we used to introduce the occupation vector $\ve N$, driven Potts models are invariant under simultaneous, discrete rotations of all oscillators. This symmetry carries over to $\ve p$ and leads to equivariance~\cite{Gol88,Cra91b,Gol03} of the macroscopic dynamics~\eqnref{eom} under the corresponding symmetry transformations, i.e.,
\algn{\eqnlab{equivariance}
	\ve h(\ve p) = \gamma^{-1}\ve h(\gamma\, \ve p)\,.
}
The transformations represented by $\gamma$ are associated with the cyclic group $C_q$~\cite{Gol88}. The group $C_q$ consists of the $q$ elements
\algn{\eqnlab{cm}
	C_q &=\{e, \rho, \rho^2,\ldots,\rho^{q-1}\}\,,
}
where $e$ denotes the identity element and $\rho$ is the generator of the group, satisfying $\rho^q=e$. In terms of $\ve p$, the symmetry induced by $C_q$ corresponds to cyclic permutations of the elements of $\ve p$:
\algn{\eqnlab{permutation}
	(p_0,p_1,\ldots,p_{q-1})^{\sf T} \overset{\rho}{\mapsto} (p_{q-1},p_0,p_1,\ldots,p_{q-2})^{\sf T}\,,
}
where the generator $\rho$ has the $q$-dimensional matrix representation
\algn{\eqnlab{rhomat}
	\mbb{R}_\rho = \pmat{
				0&1&0&\ldots&0\\0&0&1&\ldots&0\\\vdots&\vdots&\vdots&\ddots&\vdots\\
				0&0&0&\ldots&1\\1&0&0&\ldots&0	
	}\,.
}
All elements of $C_q$ in \Eqnref{cm} have a matrix representation of the kind~\eqnref{rhomat} given by $\mbb{R}_{\rho^n} = \mbb{R}_\rho^n$, where $n=1,\ldots,q$. The corresponding $q$-dimensional representation of $C_q$ is reducible, and it is decomposed into the irreducible representations of $C_q$ by means of a discrete Fourier transform, which we exploit below.

To summarise this Section, in the thermodynamic limit, the stochastic dynamics of driven Potts models becomes deterministic, and follows the macroscopic equations of motion, \Eqsref{eom}. These equations describe the evolution of the probabilities $p_n(t)$ for a randomly chosen oscillator to be in state $n$ at time $t$ and they share the microscopic symmetries of Potts oscillators. The thermodynamic observables $\langle\dot\Sigma\rangle$, $\langle\dot{S}_\text{env}\rangle$, and $\langle\dot W\rangle$, as well as the stability measure $\langle L\rangle$, that we defined for finite systems, all have deterministic, macroscopic limits, $\langle\dot\sigma\rangle$, $\langle\mc{\dot S}_\text{env}\rangle$, $\langle\mc{\dot W}\rangle$, and $\mc{L}$, respectively, that depend on $\ve p$.
\section{Stability of decoherent phase}\seclab{stabdec}
After having taken the thermodynamic limit in \Secref{thermo}, we now analyse the stability of decoherent phase in this limit.
The formulation~\eqnref{eom} of the macroscopic dynamics in terms of $j(x',x)$ immediately shows that any constant probability vector is a fixed point of the dynamics~\eqnref{eom}, because
\algn{\eqnlab{constfp}
	j(x,x) - j(x,x) = 0\,,
}
for all $x$. Probability conservation~\eqnref{constr} implies that $x=q^{-1}$, so the decoherent state $\ve p^*$ [\Eqnref{decstate}] is a fixed point of the dynamics~\eqnref{eom}. We now establish the stability and the bifurcations of the decoherent fixed point $\ve p^*$ for arbitrary $q$.

To this end, we first transform the coordinates $\ve p\to\ve x$ as
\algn{
	\ve p = \ve x+\ve p^*\,.
}
The vector $\ve x$, mentioned already in \Secref{intro}, characterises the deviations from the decoherent fixed point. In terms of $\ve x$, probability conservation~\eqnref{constr} reads $\sum_{n=0}^{q-1} x_n = 0$ and the decoherent fixed point is located at $\ve x^*=0$. The transformed macroscopic dynamics~\eqnref{eom} reads
\algn{\eqnlab{eomx}
	\ve{\dot x} = \ve h(\ve x + \ve p^*) \equiv \ve{\tilde h}(\ve x)\,,
}
which inherits the $C_q$ symmetry \eqnref{equivariance}, because $\mbb{R}_\gamma \ve p^* = \ve p^*$ for all $\gamma$ in $C_q$. 

The stability of the decoherent fixed point $\ve x^*=0$ is determined by the stability matrix
\algn{
	\mbb{M}_{nm} = \partial_m \tilde h_n(\ve x)\big|_{\ve x = 0}\,.
}
The equivariance relation~\eqnref{equivariance} then implies that $\mbb{M}$ commutes with the generator $\mbb{R}_\rho$,
\algn{\eqnlab{commu}
	[\mbb{R}_\rho,\mbb{M}] = 0\,.
}
This in turn, entails that the stability matrix $\mathbb{M}$ is circulant~\cite{Dav79}, and that it has the eigenvectors $\ve v_k$ with components
\algn{\eqnlab{evs}
	(v_k)_n = \exp\left(\frac{2\pi i k n}{q}\right)\,,
}
where $k,n=0,\ldots,q-1$. Equations~\eqnref{commu} and \eqnref{evs} are a consequence of the $C_q$ symmetry of driven Potts models but are otherwise general.

The non-zero elements of the stability matrix $\mbb{M}$ of the decoherent fixed point $\ve p^*$ read
\algn{\eqnlab{dhcomp}
	\mbb{M}_{nn} =j_{10}-j_{01}\,,\quad	\mbb{M}_{n+1 n} = j_{01}\,,\quad \mbb{M}_{n-1 n} = j_{10}\,,
}
where we denote by $j_{nm}$ the derivatives of the average flux $j$ at $\ve p^*$:
\algn{\eqnlab{fluxder}
	j_{nm} = \partial^n_{y}\partial^m_xj(y,x)|_{x=y=\frac1q}\,.
}
In order to determine the stability of the decoherent fixed point $\ve x^*$, we must diagonalise the stability matrix $\mbb{M}$.
\subsection{Discrete Fourier transform}\seclab{dft}
Because $\mbb{M}$ is circulant, a consequence of \Eqnref{commu}, it is diagonalised by Fourier transform. The discrete Fourier transform for sequences of length $q$ is expressed by a $q\times q$ matrix $\mbb{F}$. The latter is constructed from the eigenvectors~\eqnref{evs} according to

\algn{\eqnlab{fourier}
	\renewcommand*{\arraystretch}{1.3}
%	F = (v_n)_m = \exp\left(\frac{2\pi i nm}{q}\right)\,,
	\mbb{F} = \pmat{\horzbar&\ve v^{\sf T}_0&\horzbar\\
			\horzbar&\ve v^{\sf T}_1&\horzbar\\
			\horzbar&\ve{v}^{\sf T}_{2}&\horzbar\\
			&\vdots&\\
%			\horzbar&\ve v^{\sf T}_{\frac{q-1}2}&\horzbar\\
			\horzbar&\ve{v}^{\sf T}_{q-1}&\horzbar}\,,\quad\text{i.e.,}\quad
   F_{kn} = \ee^{\frac{i2\pi kn}{q}}\,.
}
The backtransform is obtained through the inverse matrix $\mbb{F}^{-1}$, given by
\algn{
	\mbb{F}^{-1} = q^{-1} \mbb{F}^\dagger,
}
where $F^\dagger_{nk} = \bar F_{kn}$ denotes Hermitian conjugation. The Fourier transform~\eqnref{fourier} maps a vector $\ve x$ to its Fourier modes $\ve{\hat x}$
\algn{\eqnlab{fourierx}
	\ve{\hat x}  = \mbb{F} \ve x\,.
}

Expressing the problem in Fourier modes takes advantage of the periodicity of the system. Since the vector $\ve x$ is real, the components of $\ve{\hat x}$ occur in complex conjugate pairs of Fourier modes $z_k$ and $\bar z_k$ with
\algn{\eqnlab{xzrel}
	\hat x_{k} = z_k\,,\qquad \hat x_{-k} = \bar z_k\,,
}
where $k=1,\ldots,\lfloor\frac{q}2\rfloor$ (indices modulo $q$). 

Note that up to two Fourier modes are real: First, we have
\algn{
 	\hat x_0 = \hat x_{-0}\,,\qquad z_0 = \bar z_0\,.
}
This component can safely be neglected, because $\hat x_0 = 0$ due to probability conservation~\eqnref{constr}. Disregarding $\hat x_0$ reduces the dimensionality of the problem from $q$ to $q-1$.

Second, for even $q$, the Fourier mode with $k=\frac{q}2$ is real, since
\algn{\eqnlab{realF}
	\hat x_{\frac{q}2} = \hat x_{-\frac{q}2}\,,\qquad z_{\frac{q}2} = \bar z_{\frac{q}2}\,.
}
This Fourier mode is associated with alternating probability patterns, as we discuss in \Secref{stabexp}. Therefore, we henceforth call $z_{\frac{q}2}$ ``alternating mode''.

Altogether, the dynamics is $q-1$ dimensional and described by the $\lfloor \frac{q}2 \rfloor$ dynamic Fourier modes $z_1,\ldots,z_{\lfloor\frac{q}2\rfloor}$, as advertised for $q=9$ in \Secref{intro}, where $\lfloor\frac{q}2\rfloor = 4$.
\subsection{Stability exponents}\seclab{stabexp}
With help of the Fourier transform~\eqnref{fourier}, we diagonalise $\mbb{M}$ as $\mbb{D} = \mbb{FMF}^{-1}$. The diagonal components $\mbb{D}_k\equiv\mbb{D}_{kk}$ of $\mbb{D}$ read
\algn{\eqnlab{eig}
	\mbb{D}_{k} = \mu_k +i\omega_k\,,\qquad \mbb{D}_{-k} = \mu_k -i\omega_k\,,
}
for $k=0,\ldots,\lfloor \frac{q}2\rfloor$. Here $\mu_k$ and $\omega_k$ are the real and imaginary parts of the stability exponent of the $k^\text{th}$ Fourier mode close to ${\ve x}^*$. From $\mbb{M}$ in \Eqnref{dhcomp}, we obtain
\algn{\eqnlab{muom}
	\mu_k =  \Lambda\sin^2\!\left(	\frac{\pi k}q\right)\,,\qquad	\omega_k = \Omega\sin\left(	\frac{2\pi k}q\right)\,,
}
where we introduced the parameters
\algn{\eqnlab{lamom}
	\Lambda = 2(j_{10}-j_{01})\,,\qquad  \Omega = j_{10}+j_{01} = qj_{00}\,.
}
\begin{figure}
	\includegraphics{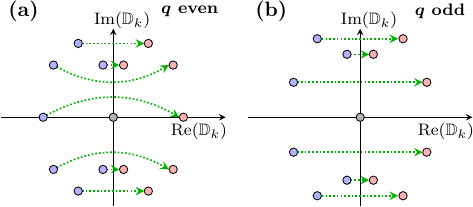}
	\caption{High-dimensional Hopf bifurcation of the decoherent phase for even and odd $q$. The real parts of all non-zero eigenvalues (blue bullets for $\Lambda<0$) change sign at $\Lambda=0$ and become positive for $\Lambda>0$ (red bullets). The eigenvalue of the zero mode is shown as the grey bullet. (a) Even $q=8$: There is a nontrivial real mode that changes sign at $\Lambda=0$. All other non-zero eigenvalues are complex. (b) Odd $q=7$: All non-zero eigenvalues are complex.}\figlab{hopf}
\end{figure}

The factor $\sin^2\!\left(\frac{\pi k}q\right)\geq0$ in  \Eqnref{muom} is non-negative and zero only for $k=0$, where $\mbb{D}_0=0$. The corresponding zero mode $\ve v^{\sf T}_0 = (1,..,1)$ is a consequence of probability conservation~\eqnref{constr} and it originates from the translational invariance of $\ve{\tilde h}$ along the unphysical, one-dimensional subspace normal to $\sum_{n=0}^{q-1} x_n = 0$.

Comparing \Eqnref{muom} to the stability exponents~\eqnref{muomsingle} of a single uncoupled oscillator, we find
\sbeqs{\eqnlab{muom2}
\algn{
	\mu_k =& 	\mu_k |_{\mathscr{J}=0} \left[ 1 - \frac{2}q \frac{{k^{+}}'(0) + {k^{-}}'(0)}{k^++k^-}\right]\,,\eqnlab{mu}\\
	\omega_k =& \omega_k|_{\ms{J}=0}\eqnlab{om}\,,
}
}
where $k^{\pm}{'}(0)$ denotes the derivative of $k^\pm(x)$ at $x=0$ and $k^\pm(0) = k^\pm$. This shows that interactions affect the stability of the decoherent phase, determined by $\ve \mu$, because the additional term in \Eqnref{mu} may change the sign of $\ve \mu$.

From \Eqnref{muom} we observe that the real parts $\mu_k$ have the same sign as $\Lambda$. This implies that the decoherent phase is stable for $\Lambda<0$ and unstable for $\Lambda>0$. We therefore call $\Lambda$ the bifurcation parameter.

The natural frequencies $\omega_k$ in \Eqnref{om}, by contrast, are unaffected by interactions at the decoherent fixed point. The fact that $\ve \omega$ remains finite for $k\neq\frac{q}2$ when $\Lambda$ changes sign indicates the emergence of oscillatory states through a high-dimensional Hopf bifurcation~\cite{Guc83,Cra91a}.

The behaviour of the stability exponents close to the bifurcation is shown schematically in \Figref{hopf}. Figure~\figref{hopf}(a) shows the behaviour for even $q$, \Figref{hopf}(b) for odd $q$. In both cases, the real parts $\ve \mu$ of all stability exponents, except for the one associated with the zero mode, cross the imaginary axis when $\Lambda$ changes sign. For odd $q$ [\Figref{hopf}(b)] all physical exponents occur in complex conjugate pairs. For even $q$ [\Figref{hopf}(a)], by contrast, there is a unique, real eigenvalue for $k = \frac{q}2$ with
\algn{\eqnlab{unstmode}
	\mbb{D}_{\frac{q}{2}} =\mu_{\frac{q}{2}} = \Lambda\,, \quad\text{and}\quad \ve v_{\frac{q}{2}} = (1,-1,\ldots,1,-1)^{\sf T}.
 }
The corresponding eigenvector $\ve v_{\frac{q}2}$ is aligned with the alternating mode $z_\frac{q}2$ in \Eqnref{realF}, and unstable for $\Lambda>0$.

The instability of the decoherent fixed point along $\ve v_{\frac{q}2}$ for even $q$ can be understood intuitively: The ferromagnetic interactions of the Potts model energetically favours oscillators to be in the same state, while the thermal noise favours states of high entropy. Even $q$ allows for a stationary, alternating pattern of $\ve x$ with higher occupations in ``every other state'' of the oscillator, as shown in \Figref{modes}(a). This alternating pattern represents a compromise between the energetic and entropic tendencies within the system for even $q$. For odd $q$, such states are frustrated, because they are incommensurate with the periodicity of the oscillators, see \Figref{modes}(b). Patterns with fewer maxima and minima, however, may form as rotating patterns due to the driving $f$.
\begin{figure}
	\includegraphics{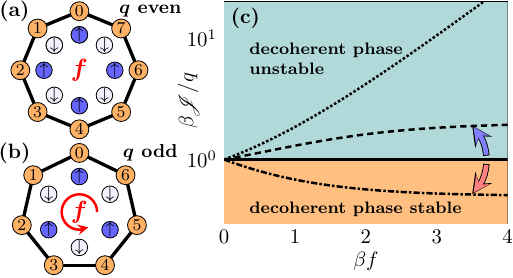}
	\caption{(a) Alternating mode for even $q=8$. A stationary, alternating pattern with increased probability in four states (blue) is commensurate with the topology of the ring. (b) No alternating mode for odd $q=7$. Alternating patterns, here with increased probability in three states (blue), are incommensurate with the topology of the ring, and rotate due to the driving $f$. (c) Stability of decoherent phase in the plane spanned by $\beta f$ and $\beta\!\!\ms{J}/q$ with phase boundaries for different dynamics (black lines). Some dynamics stabilise the decoherent phase (blue arrow) others destabilise it (red arrow): Mixed Arrhenius dynamics [\Eqnref{karr}] with $\xi = 1$ (dotted line), $\xi = 0.5$ (dashed line), and $\xi = -1$ (dash-dotted line). The solid line corresponds to classical Arrhenius dynamics ($\xi =0$) and Glauber dynamics [\Eqnref{kgla}], described in the main text.
	}\figlab{modes}
\end{figure}

The dynamics, i.e., the choice of $k^{\pm}(x)$ in the transition rates defined in \Eqnref{rates}, has a significant impact on the stability of the decoherent phase. Different choices for $k^{\pm}(x)$ lead to different parameter regions where the decoherent phase is stable, i.e., where $\Lambda>0$. Figure~\figref{modes}(c) shows the phase diagram for the stability of the decoherent phase. Here, $f=0$ corresponds to the equilibrium Potts model with global interactions, where $\ve p^*$ becomes unstable at $\beta\!\!\ms{J}_c/q = 1$~\cite{Wu82}.

For $\beta f>0$, the phase boundary between the stable and unstable regions depends on the dynamics, as shown by the different broken lines in \Figref{modes}(c). Each line corresponds to a choice of $k^{\pm}(x)$ and separates parameter regions where decoherence is stable (below the line) and unstable (above the line). As indicated by the blue arrow, some dynamics stabilise the decoherent phase compared to the equilibrium case, other destabilise it (red arrow). Interestingly, some specific dynamics conserve the phase boundary of the undriven, equilibrium model (solid line).

For instance, the rescaled, macroscopic versions $k_\text{Arr}^{\pm}(x)$ of the mixed Arrhenius rates $K_\text{Arr}^{\pm}(x)$ introduced in \Eqnref{Karr} have the form
\algn{\eqnlab{karr}
	k_\text{Arr}^{\pm}(x) = \frac1{\tau}\ee^{\frac{\beta}{2}\left[(1\mp\xi)\ms{J}x\pm f\right] }\,.
}
These rates lead to a change of sign of $\mu_k$, i.e. $\Lambda=0$, at the critical coupling
\algn{
	\frac{\beta\!\! \ms{J}_c}{q} = \frac1{1-\xi\text{tanh}\left(\frac{\beta f}{2}\right)}\,.
}
The choice $\xi=0$ corresponds to the classical Arrhenius rates, for which the driven Potts model was studied in Refs.~\cite{Her18,Her19}. This choice of dynamics is particularly convenient, since the phase boundary $\beta\!\! \ms{J}_c/q=1$ is identical to that at equilibrium and independent of $f$ [see solid line in \Figref{modes}(c)]. Positive values, $\xi>0$, push the phase boundary to larger values of $\beta\!\! \ms{J}_c/q$, stabilising the decoherent phase, while negative values, $\xi<0$, destabilise the decoherent phase.

The rescaled versions~$k_\text{Gla}^{\pm}(x)$ of the Glauber rates [\Eqnref{Kgla}] read in the thermodynamic limit
\algn{\eqnlab{kgla}
	k_\text{Gla}^{\pm}(x) = \frac2{\tau}\frac{\ee^{\pm \frac{\beta f}2 }}{\ee^{-\beta\!\!\ms{J}x}+1}\,.
}
This set of rates has similar advantages to the classical Arrhenius rates in that the critical coupling $\beta\!\! \ms{J}_c/q=1$ is identical to that at equilibrium, conserving the equilibrium phase boundary of the decoherent phase. The origin of this behaviour is understood by noticing that ${k^{\pm}}'(0) = \frac{\beta\mathscr{J}}2 k^\pm$ for both classical Arrhenius rates and for Glauber rates, and using \Eqnref{mu}. We therefore use either Arrhenius rates~\eqnref{karr} with $\xi=0$ or Glauber rates~\eqnref{kgla} in the numerical simulations presented in \Secref{numsim}.
\section{Coherent small-amplitude oscillations}\seclab{saoscil}
From the stability analysis in \Secref{stabdec}, we concluded that the decoherent fixed point becomes unstable for $\Lambda>0$, at a phase boundary that depends on the dynamics. We now establish the final states that the system evolves to in the long-time limit, for $0<\Lambda\ll1$ and $\Lambda\ll\beta f$. It turns out that the choice of dynamics not only affects the stability of the decoherent phase, but it also changes the characteristics of the coherent small-amplitude states that occur when the decoherent phase is unstable.

In order to show this, we first Fourier transform the equation of motion~\eqnref{eom}, to obtain the dynamics of the Fourier modes $\ve{\hat x}$.
\subsection{Dynamics of Fourier modes}
In terms of the Fourier modes~\eqnref{fourier}, the equation of motion~\eqnref{eomx} takes the form
\algn{\eqnlab{eomz}
	\dot{\ve{\hat x}} =  \ve{\hat h}(\ve{\hat x})\,, \qquad  \ve{\hat h}(\ve{\hat x})\equiv \mbb{F}\ve{\tilde h}(\mbb{F}^{-1}\ve{\hat x})\,.
}
Close to the bifurcation, we expect that the system remains in the vicinity of the decoherent fixed point $\ve{\hat x} = 0$ and expand $\ve{\hat h}(\ve{\hat x})$ to third order in $\ve{\hat x}$:
\algn{\eqnlab{eomexp}
	\ve{\dot{\hat{x}}} \sim \ve{\hat h}^{(1)}\!\!(\ve{\hat x}) + \ve{\hat h}^{(2)}\!\!(\ve{\hat x}) + \ve{\hat h}^{(3)}\!\!(\ve{\hat x})		\,,
}
where the partial derivatives of $\ve{\hat h}$ at $\ve{\hat x}=0$ determine the model-dependent coefficients of the expansion:
\algn{\eqnlab{mcoefs}
	\ve{\hat h}^{(1)}\!\!(\ve{\hat x}) &= \mbb{D}\ve{\hat x}\,,\quad
	\ve{\hat h}^{(2)}\!\!(\ve{\hat x}) = \frac1{2!}\sum_{k',k''=1}^{q-1}\partial_{k'}\partial_{k''}\ve{h}(\ve{\hat x})\big|_{\ve{\hat x}=0}\hat x_{k'} \hat x_{k''}\,,\nn\\
	\ve{\hat h}^{(3)}\!\!(\ve{\hat x}) &= \frac1{3!}\sum_{k',k'',k'''=1}^{q-1}\partial_{k'}\partial_{k''}\partial_{k'''} \ve{h}(\ve{\hat x})\big|_{\ve{\hat x}=0}\hat x_{k'} \hat x_{k''} \hat x_{k'''}\,.
}
Conveniently, these coefficients are constrained by the $C_q$ symmetry [\Eqnref{equivariance}], which translates to $\ve{\hat h}$ as
\algn{
	\ve{\hat h}(\ve{\hat x}) = \mbb{\hat R}_\gamma^{-1}\ve{\hat h}(\mbb{\hat R}_\gamma \ve{\hat x})\,,
}
where $\mbb{\hat R}_\gamma \equiv \mbb{F}\mbb{R}_\gamma \mbb{F}^{-1}$ with $\mbb{R}_\gamma=\mbb{R}_\rho^n$, $n=0,\ldots,q-1$, and generator $\mbb{R}_\rho$ given in \Eqnref{rhomat}.

$\mbb{R}_\rho$ is diagonalised by $\mbb{F}$, and we have $\mbb{\hat R}_\rho = \mbb{F}\mbb{R}_\rho \mbb{F}^{-1}$ with diagonal components $\mbb{\hat R}_k\equiv (\mbb{\hat R}_\rho)_{kk}$ given by
\algn{\eqnlab{rhoirrep}
	\mbb{\hat R}_{\pm k} = \ee^{\pm 2\pi i k/q}\,,
}
for $k=0,\ldots,\lfloor \frac{q}2\rfloor$. All other symmetry transformations $\mbb{\hat R}_\gamma$ are constructed from the generator $\mbb{\hat R}_\rho$ by $\mbb{\hat R}_\gamma=\mbb{\hat R}^n_\rho$, $n=0,\ldots,q-1$. 

Equation~\eqnref{rhoirrep} highlights a major advantage of the Fourier-transformed coordinates: Through $\mbb{F}$, the $q$-dimensional reducible representation of $C_q$ associated with \Eqnref{rhomat} is transformed into the direct sum of $q$ one-dimensional irreducible representations of $C_q$, given in \Eqnref{rhoirrep}. Out of these, only the trivial representation with $k=0$ is real for odd $q$. This representation is associated with the unphysical zero mode which has its origin in the constraint~\eqnref{constr}.

For even $q$, an additional real representation with generator $\mbb{\hat R}_{\frac{q}2}=-1$ is associated with the alternating mode [\Eqnref{unstmode}]. The existence of this additional real representation of the symmetry group for even $q$ is the group-theoretic explanation for the occurrence of $\ve v_{\frac{q}2}$ in~\Eqnref{unstmode}.

The decomposition~\eqnref{rhoirrep} into irreducible representations is a significant simplification, because the Fourier-transformed symmetry transformations $\mbb{\hat R}_\rho$ act locally on the components $\ve{\hat x}$, instead of globally shuffling the components of $\ve x$ as with $\mbb{R}_\rho$ in \Eqnref{rhomat}.

In particular, for the expansion \eqnref{eomexp}, the symmetry implies that $\ve{\hat h}^{(2)}$ and $\ve{\hat h}^{(3)}$ [see \Eqnref{mcoefs}] contain only terms with $k''=k-k'$ and $k''' = k-k'-k''$, respectively, which significantly simplifies the manipulations that follow.
\subsection{Normal form}
Close to the bifurcation, for $0<\Lambda\ll1$ and $\Lambda\ll\beta f$, the coupled non-linear equation~\eqnref{eomexp} can be brought into normal form~\cite{Cra91a} by a non-linear transformation $\ve {\hat x}\mapsto\ve {\hat x'}$. This normal form characterises bifurcations into small-amplitude states, i.e. synchronised states and stationary patterns. It is considerably simpler than \Eqnref{eomexp} as it contains only the essential terms for the bifurcation. In terms of the transformed components $z'_k$, the normal form reads
\algn{\eqnlab{bhnf}
	\dot{z}'_{k} \sim \bigg(\mbb{D}_{k} - \sum_{k=1}^{\lfloor \frac{q}2\rfloor}\mbb{C}_{kk'}|z'_{k'}|^2 \bigg) z'_k\,,
}
where $\mbb{C}_{kk'}$, $k,k'=1,\ldots,\lfloor \frac{q}2\rfloor$, are coefficients that determine the magnitudes of the essential bifurcation terms.

The reason why \Eqnref{bhnf} is easier to analyse than \Eqsref{eomexp} is that \Eqnref{bhnf} allows to decouple the phases and amplitudes of $z'_k$ by the coordinate transform $z'_k =r_k\ee^{i\phi_k}$. The corresponding phase-amplitude equations for
\algn{
\ve r\equiv(r_1,\ldots,r_{\lfloor\frac{q}2\rfloor})^{\sf T}\,,\quad\text{and}\quad\ve\phi\equiv(\phi_1,\ldots,\phi_{\lfloor\frac{q}2\rfloor})^{\sf T}
}
are analysed in \Secref{amp}. 

The $\lfloor\frac{q}2\rfloor\times\lfloor\frac{q}2\rfloor$ coefficient matrix $\mbb{C}$ in \Eqnref{bhnf} decomposes into its real and imaginary parts as
\algn{\eqnlab{Ceqn}
	\mbb{C} = \mbb{A} + i \mbb{B}\,,
}
where $\mbb{A}$ and $\mbb{B}$ are real. We obtain $\mbb{C}$ from a method outlined in \Secref{bhnf}.

We note that in normal form~\eqnref{bhnf} the expanded equations of motion~\eqnref{eomexp} appear to have a higher degree of symmetry: While \Eqnref{eomexp} is equivariant under discrete rotations with $\mbb{\hat R}_\rho$ [\Eqnref{rhoirrep}], the normal form is equivariant under continuous rotations of the phases of $z'_k$.  However, the transformation that brings \Eqnref{eomexp} into the normal form~\eqnref{bhnf} generates higher-order terms that we neglect. These higher-order terms are equivariant under the discrete complex rotations~\eqnref{rhoirrep}, but they are inessential for the local dynamics close to the bifurcation.

Hence, the terms of lower symmetry have not actually disappeared from \Eqnref{bhnf} but are merely transformed into the higher orders, leaving us with only the essential terms of the Hopf bifurcation.
\subsection{Transformation into normal form}\seclab{bhnf}
We now outline the method to bring the expanded macroscopic equation of motion~\eqnref{eomexp} into the normal form~\eqnref{bhnf}. To third order in $\ve{\hat x}$ this is achieved by the non-linear coordinate transformation $\ve{\hat x}\mapsto\ve{\hat x}'$, where
\algn{\eqnlab{trafo}
	\ve{\hat x'} = \ve{\hat x} + \ve{f}^{(2)}(\ve{\hat x}) + \ve{f}^{(3)}(\ve{\hat x})\,.
}
Here, $\ve{f}^{(n)}(\ve{\hat x})$ contains polynomials of order $n=2,3$ in $\ve{\hat x}$. Since $\ve{\hat x'}\sim \ve{\hat x}$ to linear order in $\ve{\hat x}$, the transformation does not affect the linear order in \Eqnref{eomexp}. Instead, the transformation can be chosen to eliminate the non-linear cross terms in \Eqnref{eomexp} and to bring the dynamics in the normal form~\eqnref{bhnf}~\cite{Elp87}. 

As we explain below, we need only find the quadratic transformation $\ve{f}^{(2)}$ explicitly. The explict form of the quartic transformation $\ve{f}^{(3)}$ is not required.

In Appendix~\secref{appctrafo} we show that $\ve{f}^{(2)}$ is given by the condition~\cite{Guc83,Cra91a} 
\algn{\eqnlab{L2eqn}
	\mbb{L}(\ve f^{(2)}) = \ve{\hat h}^{(2)}\,, \quad \mbb{L}(\ve f)\equiv 	 (\ve f\cdot\nabla) \ve{\hat h}^{(1)}  - \ve{\hat h}^{(1)\sf T} D\ve f\,,
}
where $(D\ve{f})_{kk'} = \partial f_k/\partial \hat x_{k'}$ denotes the matrix of derivatives of a given vector valued function $\ve f$. $\mbb{L}$ is a linear mapping that maps the unknown function $\ve f^{(2)}$ to the known function $\ve{\hat h}^{(2)}$.

To determine $\ve f^{(2)}$, we compute the inverse of $\mbb{L}$ and apply it to $\ve{\hat h}^{(2)}$:
\algn{
	\ve f^{(2)} =  \mbb{L}^{-1}(\ve{\hat h}^{(2)})\,.
}
A convenient matrix representation for $\mbb{L}$ is obtained by expressing $\ve f^{(2)}$ and $\ve{\hat h}^{(2)}$ in a basis in which $\mbb{L}$ is diagonal~\cite{Cra91a}. Such a basis is given by the elements $\ve{e}_{k'}$ with components
\algn{\eqnlab{basel}
	 (\ve{e}_{k'})_k = e_{k'k} \equiv \css{		{\hat x}_{k'}{\hat x}_{k-k'}\,, & k\neq k'\\
	 					0\,,	& k=k'
					}\,,
}
where $k,k'= 1,\ldots q-1$. Substituting $e_{k'k}$ into $\mbb{L}$, we find
\algn{
	\mbb{L}_k(\ve{e}_{k'}) = \left[\mu_k -\mu_{k'}-\mu_{k-k'} + i\left(\omega_k-\omega_{k'}-\omega_{k-k'}\right)\right]e_{k'k}\,.
}
Hence, in a basis formed by the elements in \Eqsref{basel}, $\mbb{L}$ is diagonal and its inverse can be computed as
\algn{\eqnlab{invL}
	\mbb{L}^{-1}_k(\ve{e}_{k'}) = \frac{e_{k'k}}{\mu_k -\mu_{k'}-\mu_{k-k'} + i\left(\omega_k-\omega_{k'}-\omega_{k-k'}\right)}\,.
}
As shown in Appendix~\secref{gender}, $\ve{h}^{(2)}$ reads in terms of the elements~\eqnref{basel},
\algn{
	\hat h_k^{(2)} = \sum_{k'=1}^{q-1} \hat H^{(2)}_{kk'} e_{k'k}\,,
}
where
\algn{
	\hat H^{(2)}_{kk'} = \frac1q\frac{j_{20}-j_{02}}{\Lambda}\mu_{k} - i\frac{j_{11}}{q^2j_{00}}\left(\omega_k-\omega_{k'} -\omega_{k-k'}\right)\,.
}
Given this, we can immediately compute $\ve{f}^{(2)}$ using \Eqnref{invL}:
\algn{
	{\hat f}_k^{(2)} = \sum_{k'=1}^{q-1} \frac{\hat H^{(2)}_{kk'}e_{k'k}}{\mu_k -\mu_{k'}-\mu_{k-k'} + i\left(\omega_k-\omega_{k'}-\omega_{k-k'}\right)}\,,
}
which provides the transformation required to eliminate the quadratic terms in \Eqnref{eomexp}.

In addition to removing $\ve{\hat h}^{(2)}$, the quadratic transformation $\ve f^{(2)}$ also impacts the cubic and higher-order terms in \Eqnref{eomexp}. We show in Appendix~\secref{appctrafo} that $\ve f^{(2)}$ alters the third-order terms as
\algn{\eqnlab{g3eqn}
	\ve{\hat h}^{(3)}\mapsto \ve{\hat g}^{(3)} \equiv \ve{\hat h}^{(3)} + \ve{\hat h}^{(2)\sf T}D\ve f^{(2)}\,.
}

The role of the cubic transformation $\ve f^{(3)}$ in \Eqnref{trafo}, in turn, is to remove all inessential third-order terms contained in $\ve{\hat h}^{(3)}$ and those generated by $\ve f^{(2)}$. $\ve f^{(3)}$ is determined in a similar way as before, by constructing the inverse of the mapping $\mbb{L}$~\cite{Cra91a}. However, since we know that the cubic transformation $\ve f^{(3)}$ removes the non-essential third-order terms, but otherwise affects only the terms of order four and higher, we do not need to determine $\ve f^{(3)}$ explicitly. Instead, we only need to compute the effect of $\ve f^{(2)}$ on the coefficients of the \textit{essential} cubic terms in \Eqnref{bhnf}.

Hence, once we have computed $\ve{\hat g}^{(3)}$ in \Eqnref{g3eqn}, the coefficients $\mbb{C}$ in the normal form~\eqnref{bhnf} are determined by projecting $\ve{\hat g}^{(3)}$ onto the third-order terms given in \Eqnref{bhnf}. This procedure brings the original equations~\eqnref{eomexp} into normal form~\eqnref{bhnf} with coefficients $\mbb{D}$ and $\mbb{C}$. The details of the projection of $\ve{\hat g}^{(3)}$ and how it leads us to $\mbb{C}$ are summarised in Appendix~\secref{gender}.
\subsection{Amplitude equations}\seclab{amp}
Once we have obtained the normal form~\eqnref{bhnf}, the equations for the amplitudes $\ve r$ and the phases $\ve \phi$ follow by applying the transformation $z'_k = r_k\ee^{i\phi_k}$. The amplitudes obey the equations
\sbeqs{
\algn{\eqnlab{amp}
	\dot r_k \sim \big(\mu_k - \sum_{k'=1}^{\lfloor \frac{q}2\rfloor}\mbb{A}_{kk'}r^2_{k'}\big)r_k\,,
}
with matrix $\mbb{A}$ [cf. \Eqnref{Ceqn}] given by
\algn{\eqnlab{amat1}
	\mbb{A} = A\left(\ve u \ve v^{\sf T} - \mbb{W}\right)\,,\quad A = \frac{4}{q^2}(j_{03}-j_{30})\,,
}
where $\ve u$, $\ve v$ are vectors and $\mbb{W}$ is a diagonal matrix with components
\algn{\eqnlab{amat2}
	u_k &= 1+a\cos\left(\frac{2\pi k}{q}\right)\,,\quad v_{k'} = \left(1-\frac12\delta_{k'\frac{q}{2}}\right)\sin^2\left(\frac{\pi k'}{q}\right)\,,\nn \\
	\mbb{W}_{kk'} &= W_k\delta_{kk'}\,,\quad W_k=\frac{1}{4}\left(1+\frac13\delta_{k\frac{q}2}\right)\left[2u_k-(a+1)\right]v_k \,,
}
}
and parameter
\algn{\eqnlab{aparam}
	a = \frac{j_{02}^2-j_{20}^2}{q j_{00}(j_{03}-j_{30})} - 1\,.
}
For Arrhenius and Glauber dynamics given in \Eqsref{karr} and \eqnref{kgla}, the parameter $A$ reads $A=2\cosh(\beta f/2)$, while $a=0$ for Arrhenius and $a=1$ for Glauber dynamics.

For the analysis of the amplitude equations, we focus on stable fixed points $\ve r^*$ of \Eqnref{amp}, for which $\ve{\dot r}|_{\ve r^*}=0$. The fixed points of \Eqnref{amp} encompass both stationary patterns (with $\ve{\dot \phi}=0$) and oscillating (synchronised) solutions ($\ve{\dot \phi}\neq0$) of the original \Eqsref{eomexp}. In particular, we note that the alternating mode $z_{\frac{q}2}$ is real, which implies that $\phi_{\frac{q}2} = \dot\phi_{\frac{q}2}=0$, i.e. $z_{\frac{q}2}$ is non-oscillating. Therefore, fixed points $\ve r^*$ for which $k=\frac{q}2$ is the only mode with non-vanishing amplitude correspond to stationary, alternating probability patterns.

Equation~\eqnref{amp} is non-linear and therefore has in general multiple fixed points $\ve r^*$, which we call ``states''. Each state contains $\lfloor\frac{q}2\rfloor$ amplitudes $r^*_k$. For a given state, we call the Fourier modes with vanishing amplitude, i.e. $r^*_k=0$, ``inactive'' modes and denote the set of inactive modes of a state by $S_{ia}$. Similarly, $k$ values with $r^*_k>0$ correspond to ``active'' Fourier modes of a state, and we denote by $S_a$ the set of such active modes. Furthermore, we denote the number of active modes of a given state by $|S_a|$ and the number of inactive modes by $|S_{ia}| = \lfloor\frac{q}2\rfloor-|S_a|$.

By inspection of \Eqnref{amp}, we note that $r_k^*=0$ leads to $\dot r_k = 0$ for any state and for arbitrary combinations of $k$s. The amplitudes of the remaining, active modes can be computed explicitly, due to the simple structure~\eqnref{amat1} of the matrix $\mbb{A}$. 

To ease the notation, we use the subscripts $a$ and $ia$ for vectors and matrices restricted to the subsets $S_a$ and $S_{ia}$, respectively. With this notation we write the states $\ve r^*$ as decompositions of
\algn{\eqnlab{rdec}
	\ve r^*_{ia} = 0\,, \qquad \ve r^{*2}_{a} = \mbb{A}^{-1}_a \ve \mu_a\,.
}
Here, $\boldsymbol{r}^{*2}_a$ denotes the component-wise square of the amplitude vector $\boldsymbol{r}^{*}$, restricted to the subset $S_a$ of active modes. To compute $\ve r_a^*$ we thus need to invert $\mbb{A}_a$. Since the matrix $\mbb{A}_a$ is a sum of a diagonal part and a rank-one perturbation, see \Eqnref{amat1}, it can be inverted using the Sherman-Morrison formula~\cite{She50}:
\algn{
	\mbb{A}_a^{-1} = -\frac1{A}\left(\mbb{W}_a^{-1} - \frac{\mbb{W}_a^{-1}\ve u_a \ve v_a^{\sf T}\mbb{W}_a^{-1}}{\ve v_a^{\sf T}\mbb{W}_a^{-1}\ve u_a-1}\right)\,.
}
With this, we obtain the following expression for $\ve r^{*2}_{a}$:
\algn{\eqnlab{reqn}
	\ve r^{*2}_{a} = \lambda \mbb{W}_a^{-1}\left(\alpha_a\ve u_a - \ve {\tilde \mu}_{a} \right)\,,
}
where $\lambda \equiv \Lambda/A$ denotes the reduced bifurcation parameter and
\algn{
	\alpha_a = \frac{\ve v_a^{\sf T}\mbb{W}_a^{-1}\ve {\tilde \mu}_a}{\ve v_a^{\sf T}\mbb{W}_a^{-1}\ve u_a-1}\,,\quad\ve{\tilde \mu}_a \equiv \frac{\ve \mu_a}{\Lambda}\,.
}
The quantity $\alpha_a>0$ is a scalar that depends on the set $S_a$ of active Fourier modes and on $a$.

For a set $S_a$ to be admissible, we require that all components of $\ve r_a^{*2}$ are positive. Furthermore, to ensure that $\ve r^*$ decomposed of $\ve r_{ia}^*$ and $\ve r_a^*$ [see \Eqnref{rdec}] is approached in the long-time limit, we demand that $\ve r^*$ is a stable fixed point of \Eqnref{amp}. Stability holds if all eigenvalues of the stability matrix $\mbb{N}_{lm}\equiv\partial \dot r_l/\partial r_m$ are negative.

Using \Eqnref{amp} and the form \eqnref{amat1} of $\mbb{A}$, $\mbb{N}$ decomposes into
\algn{\eqnlab{stabmat}
	\mbb{N}_{ia} = -\Lambda\,\text{diag}(\alpha_a\ve u_{ia} - \ve{\tilde \mu}_{ia})\,,\quad \mbb{N}_{a} = -(\ve{\tilde u}_a \ve{\tilde v}_a^{\sf T} - \mbb{\tilde W}_a)\,,
}
where $\ve{\tilde u}$ and $\ve{\tilde v}$ are vectors and $\mbb{\tilde W}$ a diagonal matrix with components
\algn{
	&\tilde u_k = \sqrt{2A}r^*_k u_k\,,\quad \tilde v_{k'} = \sqrt{2A}r^*_{k'} v_{k'}\,,\nn\\
    &\tilde W_{kk'} = 2\Lambda(\alpha_a u_k -	{\tilde \mu}_k)\delta_{kk'}\,.
}
All other components of $\mbb{N}$ vanish. Hence, by permuting entries, we can bring $\mbb{N}$ into block-diagonal form where $\mbb{N}_{a}$ forms a connected block, and $\mbb{N}_{ia}$ forms a connected diagonal part. The eigenvalues of $\mbb{N}$ are then given by the eigenvalues of $\mbb{N}_{a}$ and by the diagonal elements of $\mbb{N}_{ia}$.
\begin{figure}
	\includegraphics[width=\linewidth]{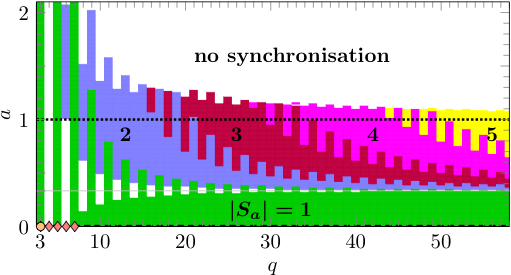}
	\caption{Phase diagram of the driven Potts model as function of $q$ and the model parameter $a$. Numbers indicate how many Fourier modes are active. The dashed line at $a=0$ denotes the location of Arrhenius dynamics, the dotted line shows $a=1$ (Glauber dynamics). The light grey line indicates $a=\frac13$, below which the problem is treated exactly in \Secref{exact13}. Symbols indicate the versions of the model considered numerically in Refs.~\cite{Her18} (bullet) and \cite{Her19} (diamonds).
	}\figlab{phasediag}
\end{figure}

To find all stable steady states, we first determine the admissible sets of active modes $S_a$ using \Eqnref{reqn}. We then diagonalise the non-trivial block $\mbb{N}_{a}$ of the stability matrix $\mbb{N}$ in \Eqnref{stabmat} to determine the regions in $q$-$a$ space, where stable coherent small-amplitude oscillations with different numbers $|S_a|$ of active modes exist. The diagonalisation of $\mbb{N}_{a}$ can be done analytically for small $|S_a|$, but numerically efficiently for higher values of $|S_a|$. 

The result of this analysis is shown as the phase diagram in \Figref{phasediag}. The white regions are associated with parameters for which no small-amplitude synchronised states exist. The remaining, differently-coloured regions host at least one stable synchronised state, where the number $|S_a|$ in the diagram indicates how many Fourier modes are active.

The white patches between the green ($|S_a|=1$) and blue ($|S_a|=2$) regions regions in \Figref{phasediag} occur for even $q$. They host no stable oscillations but stationary, alternating states with $S_a=\{\frac{q}2\}$ and $\ve{\dot \phi}=0$. The connected white region above $a>1$ does not host any stable small-amplitude states.

Figure~\figref{phasediag} shows that in the coloured regions only small numbers of Fourier modes are active in the long-time limit. The active modes are those with largest $k$.  In other words, while the active $k$s are equal or close to the maximum possible value $\lfloor \frac{q}2\rfloor$, the number $|S_a|$ of active modes is much smaller than $\lfloor \frac{q}2\rfloor$.

This implies that Potts oscillators synchronise into non-local rotating probability distributions of phases with (close to) as many minima and maxima as allowed for any given $q$  [see \Figref{oscillations}(a)].

The extreme case of this ``Potts-like'' behaviour is realised in the green parameter region ($|S_a|=1$) in \Figref{phasediag}, where only a single Fourier mode is active in any stable state. The $|S_a|=1$ region is mainly localised at $a<\frac13$, but it extends beyond $a<\frac13$ for odd $q$. We indeed show explicitly in \Secref{exact13} that for $a<\frac13$ only states with at most one active mode are stable.

Interestingly, driven Potts models exhibit multistability in the $|S_a|=1$ region; i.e., several single-mode states with different active Fourier modes are simultaneously stable, in general those modes with largest $k$. The number of stable single-mode ($|S_a|=1$) states constitutes the ``fine structure'' of the $|S_a|=1$ phase, shown in \Figref{phasediag_2}, and discussed in more detail in \Secref{exact13}. Which of these stable states the system selects in the long-time limit is discussed in \Secref{sssel}.

The calculations summarised in \Figref{phasediag} also show that regions with $|S_a|=2,3,\ldots$ occur for $a>\frac13$. We find that these regions host unique stable synchronised states for which only the $|S_a|$ Fourier modes with largest $k$ are active. The macroscopic motion for these synchronised states is quasiperiodic because the ratios of their frequencies are not in general rational. As a consequence, the trajectories of the $|S_a|$ largest-$k$ modes $z_k$ can be thought of as covering the $|S_a|$-dimensional surface of a torus.

For Arrhenius dynamics, the parameter $a$ vanishes, i.e. $a=0<\frac13$, see dashed line in \Figref{phasediag}, so the phase diagram predicts synchronised states with a single active Fourier mode for odd $q$, which is indeed what we observe in \Figsref{oscillations}(j)--(m). Figure~\figref{phasediag_2} shows that there is a second oscillating state for $q=9$, which is also stable in the thermodynamic limit. We confirm the existence of this state numerically in \Secref{numsim}.

Glauber dynamics have $a=1>\frac13$, so for increasing $q$ coherent small-amplitude oscillations with increasing number of active Fourier modes are predicted to be stable, see dotted line in \Figref{phasediag}. This prediction is confirmed numerically for $q=11$ in \Secref{numsim}.

\begin{figure}
	\includegraphics[width=\linewidth]{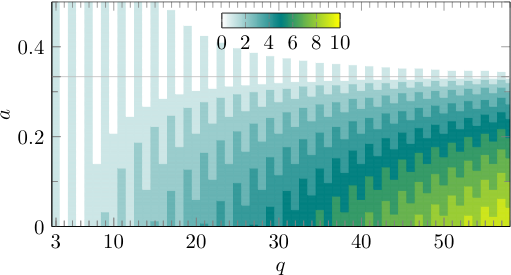}
	\caption{Fine structure of multistability in the $|S_a|=1$ phase. The colour indicates the multiplicity of stable synchronised states with a single active Fourier mode. The fine grey line shows $a=\frac13$.}\figlab{phasediag_2}
\end{figure}
\subsubsection{Analytic treatment for $a<\frac13$}\seclab{exact13}
We outline how to treat the problem exactly for $a<\frac13$, which includes Arrhenius rates with $a=0$. For $a<\frac13$, all entries of $\ve{\tilde u}_a$, $\ve{\tilde v}_a$ and $\mbb{\tilde W}$ in \Eqnref{stabmat} are positive. For this case, we show in Appendix~\secref{appamp} that by applying the matrix determinant lemma~\cite{Bro20} combined with an eigenvalue interlacing theorem~\cite{Bun78,Tho76,Wil65} one finds that the stability matrix $\mbb{N}$ has at least $|S_a|-1$ positive, i.e. unstable, eigenvalues. This implies that for $a<\frac13$, only states with a single active Fourier mode ($|S_a|=1$) are stable and states with $|S_a|>1$ are unstable. Hence, the only relevant stable small-amplitude states, synchronised and stationary, are those for which $r^*_{k}>0$ for a single $k=1,\ldots,\lfloor \frac{q}2\rfloor$ and $r^*_{k'}=0$ for all other $k'\neq k$.

For $|S_a|=1$ states, the general amplitudes given in \Eqnref{reqn} simplify to
\algn{\eqnlab{rk2}
	(r^*_{k'})^2 =\frac{ \lambda\tilde\mu_k}{u_kv_k-W_{k}}\delta_{kk'}.
}
For Arrhenius rates with $a=0$ the expression simplifies further and we obtain
\algn{\eqnlab{rk2Arrh}
	(r^*_{k'})^2 = \frac{4\lambda}3\left(1+\frac54\delta_{k\frac{q}2}\right)\delta_{kk'}\,.
}
Recall that $k=\frac{q}2$ corresponds to the stationary alternating state, present only for even $q$. For Arrhenius rates, the amplitudes of the synchronised states ($k\neq\frac{q}2$) in \Eqnref{rk2Arrh} are independent of $k$, i.e., all small-amplitude synchronised states have the same amplitude.

The stability analysis for states with one active Fourier mode and $a<\frac13$ reveals that the alternating mode ($k=\frac{q}2$) is stable for all even $q$ while the oscillating mode with largest $k$, $k=\frac{q-1}2$, is stable for all odd $q$. The remaining synchronised states with a single active Fourier mode $k$ are stable only if
\algn{\eqnlab{purestab}
	\frac34\sin^2\!\!\left(k' \pi/q\right) < \frac{\sin^2\!\!\left(k \pi/q\right)}{1+\frac43\frac{a}{a+1}\sin^2\!\!\left(k \pi/q\right)}\,,
}
for all $k'\neq k$.

Equation~\eqnref{purestab} implies that for any given $q$, the modes with large $k$, i.e., with many minima and maxima distributed over the states $0,\ldots,q-1$ have in general more stable directions. For these modes, the probability is most evenly distributed among the states $0,\ldots,q-1$, while still allowing for synchronised patterns. The number of stable states with a single active mode for given $q$ and $a$ thus depends on the number of $k=0,\ldots,q-1$ that satisfy \Eqnref{purestab} for all $k'\neq k$.

Figure~\figref{phasediag_2} shows the number of stable states with a single active Fourier mode as function of $q$ and $a$ as determined from \Eqnref{purestab}. We observe that the number of stable states with a single active Fourier mode increases as $q$ increases, and decreases as $a$ increases. Which one of these single-mode states the system evolves to at long times is discussed in \Secref{sssel}.

For even $q$, the stationary alternating state with $S_a = \{\frac{q}2\}$ is always stable. %Setting $k=\frac{q-2}2$ in \Eqnref{purestab} and $k'=1$, we find that
The first stable oscillating state for even $q$ ($S_a=\{\frac{q}2-1\}$) requires
\algn{\eqnlab{osccond}
	\sin^2\left(\frac{\pi}{q}\right)<\frac14(1-3a)\,,
}
Assuming that $a$ is non-negative, this condition can only be satisfied for $q\geq8$ and $a<\frac13$. For $q=6$ and $a=0$ the synchronised state has a marginal direction along $k'=\frac{q}2$ where the stability exponent vanishes, meaning that it is neither linearly stable nor unstable.

Equation~\eqnref{osccond} explains why no stable oscillations were observed numerically for even $q$ and Arrhenius rates ($a=0$) in Ref.~\cite{Her19}: Stable oscillations for even $q$ occur for the first time at $q=8$, while the numerical analysis in Ref.~\cite{Her19} explored only $q$ values up to $q=7$.
\subsection{Phase equations}
We now briefly analyse the phase equations. They determine the angular velocities $\ve{\dot\phi}$ with which the active modes rotate in the complex plane. The phase equations read
\sbeqs{
\algn{
	\dot \phi_k \sim \omega_k - \sum_{k'\in S_a}\mbb{B}_{kk'}r^{*2}_{k'}\,,
}
Here, the matrix $\mbb{B}$ has the entries

\algn{\eqnlab{Bmat}
	\mbb{B}_{kk'} = B\left(1-\frac12\delta_{kk'}\right)w_{k}y_{k'}\,,\quad B = \frac{1}{q^2}\frac{(j_{20}-j_{02})^2}{qj_{00}}\,,
}
with vectors $\ve w$ and $\ve y$ given by
\algn{
	w_k = \sin\left(\frac{2\pi k}q\right)\,,\quad y_{k'} = \left(1-\frac12\delta_{k'\frac{q}2}\right)\left[1+b\sin^2\left(\frac{\pi k'}q\right)\right]\,,
}
}
and parameter
\algn{\eqnlab{Bparams}
	b =4\frac{(j_{20}+j_{02})^2-\frac13(j_{30}+j_{03})qj_{00}}{(j_{20}-j_{02})^2}\,.
}
For Arrhenius and Glauber dynamics, we have $B=2\cosh^2\left(\beta f/2\right)/\sinh\left(\beta f/2\right)$, as well as $b=0$ for Arrhenius and $b=4 \tanh ^2\left(\beta f/2\right)$ for Glauber dynamics.

For $k\in S_a$ and $k\neq\frac{q}2$ we define the relative difference $\Delta \dot \phi_k$ of the oscillation frequency compared to the natural frequency $\omega_k$ as
\algn{\eqnlab{relphase}
	\Delta \dot \phi_k \equiv \frac{\dot\phi_k-\omega_k}{|\omega_k|}=-\frac{B}{\Omega}\left(\ve{y}_a\cdot\ve{r}_a^{*2} -\frac12 y_k r_k^{*2}\right)\,.
}
The right-hand side of \Eqnref{relphase} is always negative. This predicts that coherent oscillations of the active modes are consistently slower than their natural frequency $\omega_k$, with a correction of order $0<\Lambda\ll1$.
\subsection{Thermodynamic observables in synchronised phase}\seclab{obsosc}
To analyse the thermodynamics in small-amplitute states, we expand the edge functions $\mc{\dot O}$ in \Eqnref{compobs} around the decoherent fixed point $\ve x^*=0$:
\algn{
	\mc{\dot O}(y,x) \sim \mc{\dot O}_{00} + x \mc{\dot O}_{01} + y\mc{\dot O}_{10} + \frac12 x^2 \mc{\dot O}_{02} + \frac12 y^2\mc{\dot O}_{20} + xy\mc{\dot O}_{11}\,,
}
where $\mc{\dot O}_{nm} = \partial^n_y\partial^m_x\mc{\dot O}(y,x)\big|_{x=y=\frac1q}$. Substituting this expansion into \Eqnref{compobs} and Fourier transforming $\ve x\mapsto \ve{\hat x}$, we show in Appendix~\secref{appthermo} that for $0<\Lambda\ll1$, \Eqnref{obsform} takes the form
\algn{\eqnlab{obseq}
	\langle \mc{\dot O}\rangle \sim \langle \mc{\dot O}\rangle_0  -\frac{4j_{11}}{q}\ve{v}_a\cdot\ve{r^{*2}}_a\,.
}
Here, $\langle \mc{\dot O}\rangle_0 = q\mc{\dot O}_{00}$ is the average rate of change of the observable $\mc{O}$ for a single, uncoupled oscillator. We note that $\langle \mc{\dot O}\rangle$ is time-independent, so that $\llangle\mc{\dot O}\rrangle \sim \langle\mc{\dot O}\rangle$ to leading order in $\Lambda\ll1$. For the relative change $\Delta\mc{\dot O}$ at the bifurcation, we find
\algn{\eqnlab{dO}
	\Delta\mc{\dot O}\equiv\frac{\langle \mc{\dot O}\rangle - \langle \mc{\dot O}\rangle_0}{|\langle \mc{\dot O}\rangle_0|} \sim& -2\Gamma \lambda \alpha_a\,, \quad \Gamma = \frac{2j_{11}}{q^2 j_{00}}>0\,,
}
where $\Gamma$ is a dynamics-dependent parameter. For example, $\Gamma$ takes the values $\Gamma=\frac12$ and $\Gamma=1$ for Arrhenius [\Eqnref{karr}] and Glauber [\Eqnref{kgla}] dynamics, respectively.

Equation~\eqnref{dO} shows that $\Delta\mc{\dot O}$ is negative for all thermodynamic observables $\langle\dot \sigma\rangle$, $\langle\mc{\dot S}_\text{env}\rangle$, and $\langle\mc{\dot W}\rangle$. Hence, dissipation is reduced in small-amplitude states, synchronised and stationary, for all dynamics and all $q$. Equation~\eqnref{dO} generalises the observations made in \Figref{oscillations}(k) to all dynamics and all $q$.

Whether synchronisation increases or decreases dissipation in coupled oscillators is a debated question in the literature. Zhang \textit{et al.}~\cite{Zha20} argue that synchronised states in a model with global, chemical interactions, modelled by exchange reactions between oscillator phases, dissipate more energy than if they oscillate decoherently. In their model, interactions between oscillators occur at a given rate, independently of the individual rotating motion of the oscillators. As a result, alignment, and thus synchronisation, requires additional energy.

In driven Potts models, by contrast, synchronisation reduces dissipation, because collective oscillatory motion of Potts spins reduces the rate of energy-barrier crossings in the global potential, compared to decoherent oscillations.

Equation~\eqnref{dO} also implies that the derivative of the average entropy production rate $\langle\dot\sigma\rangle$  with respect to the (reduced) bifurcation parameter $\lambda$ has a finite jump of size $-2\Gamma\alpha_a\langle\dot\sigma\rangle_0$ at $\lambda=0$. Similar discontinuities have been observed numerically in thermodynamically consistent chemical reaction models across synchronisation transitions~\cite{Ngu18}.
\subsection{Stability of coherent oscillations}
Finally, we analyse the phase-space contraction rate $\mc{L}$
\algn{\eqnlab{pscr}
	 \mc{L} = -\sum_{n=0}^{q-1}\frac{\partial \dot p_n}{\partial p_n}\,,
}
close to the transition. In Appendix~\secref{apppscontr} we show that $\mc{L}$ takes the following form:
\algn{\eqnlab{L}
	\mc{L} \sim& -\frac{q\Lambda}2 + q A\,\ve{v}_a\cdot\ve{r^{*2}}_a = -\frac{q\Lambda}{2} + q\Lambda \alpha_a\,,
}
when $0<\Lambda\ll1$, independent of time $t$. With the phase-space contraction rate at the decoherent fixed point $\mc{L}_0=-q\Lambda/2$, we then find for the relative change $\Delta\mc{L}$ of the $\mc{L}$ at the bifurcation
\algn{\eqnlab{dL}
	\Delta\mc{L} \equiv \frac{\mc{L}-\mc{L}_0}{|\mc{L}_0|} = 2\alpha_a>0\,.
}
Hence, phase-space contraction increases upon crossing the phase transition. Combining this relation with \Eqnref{dO} for $\mc{\dot O}=\dot\sigma$ we find the linear relation
\algn{\eqnlab{sdr}
	\Delta\dot \sigma \sim -\Gamma \lambda \Delta\mc{L}\,.
}

This stability-dissipation relation, the main subject of the accompanying Ref.~\cite{Mei24a}, states that the relative amount by which dissipation is reduced in small-amplitude states is proportional to the relative change in phase-space contraction rate $\Delta\mc{L}$. The prefactor is given by the positive constants $\lambda$ and $\Gamma$, which depend on the dynamics through the derivatives of the average flux $j$ at the decoherent fixed point. Although the values of these constants are dynamics dependent, relation~\eqnref{sdr} holds in general and is, in particular, valid arbitrarily far from equilibrium.

Previous works on the connection between stability and dissipation either refer to near-equilibrium situations~\cite{Pri71,Kon14} or employ definitions of entropy production without clear thermodynamic interpretation~\cite{Dae99}. For so-called thermostated, deterministic systems, the phase space contraction rate can be shown to coincide with the entropy production rate in the macroscopic limit~\cite{Rue96,Sea00,Eva07}. However, the thermodynamics of these systems is quite intransparent, in particular with respect to the physical interpretation of the deterministic thermostat. The dissipation-stability relation~\eqnref{sdr}, by contrast, holds far from equilibrium, for thermodynamically consistent stochastic systems which carry a clear thermodynamic interpretation. Relation~\eqnref{sdr} shows that entropy production and phase-space contraction are not equal in stochastic systems with local detailed balance, and obey a linear relation only close to the decoherent state.
\subsection{Implications for steady-state selection}\seclab{sssel}
We have shown in \Eqnref{sdr} that close to the synchronisation transition, dissipation is minimal in small-amplitude states with highest phase-space contraction rate. In the vicinity of this transition and in the presence of multistability, this suggests that driven Potts models favour, i.e. select, states that minimise dissipation, in accordance with a minimum-dissipation principle~\cite{Pri55} that holds far from equilibrium.

In the strict thermodynamic limit, where fluctuations are negligible, however, maximum stability alone does not guarantee that the most stable state is assumed by the system. In addition, the initial condition must lie in its basin of attraction. At large but finite $N$, however, finite-size fluctuations induce transitions between nearby non-equilibrium states, thus allowing all states to be visited in the long-time limit.

For well-separated stable states at large $N$, such noise-induced transitions occur as large deviations and are exponentially suppressed in $N$~\cite{Fal23}. Close to the transition ($\lambda\ll 1$), however, the decoherent state and all stable small-amplitude states are only separated by distances $\sim\sqrt{\lambda}$ and typical (Gaussian) fluctuations around these states are roughly of the order $\sim1/\sqrt{\mc{L} N}$. With $\mc{L}\sim \Lambda\sim \lambda$ [see \Eqnref{L}], we thus find that for $1\ll N\lesssim \lambda^{-2}$, noise-induced transitions between the stable small-amplitude states occur frequently, due to typical fluctuations around the deterministic dynamics.

In this case, all stable small-amplitude states can be accessed from the decoherent state and are visited regularly. As a consequence, the states with smallest dissipation are occupied most of the time, because they have the largest average inflow rate $\langle L\rangle$. In other words, once a (stochastic) trajectory has reached a state of low dissipation, it typically spends a long time there, compared to the average time it took the trajectory to enter that state. Hence, close to the transition, at large but finite systems size, driven Potts models are expected to probabilistically select, i.e. predominantly occupy, those small-amplitude states that dissipate the least entropy.
\section{Numerical simulations}\seclab{numsim}
We now compare our results with numerical simulations of the stochastic dynamics. To this end, we generate a large number of trajectories $\ve N(t)$ that we evolve with the Gillespie algorithm~\cite{Gil76}. Along each trajectory, we compute the amplitudes $\ve r^*$ and the associated oscillation frequencies $\ve{\dot \phi}^*$, the average dissipation $\llangle\mc{\dot W}\rrangle =\llangle\dot\sigma\rrangle$, and the average phase-space contraction rate $\mc{L}$. We evolve the trajectory of a large system ($N=10^6$) for a long time $T\approx 10^2\tau\gg T_p$, where $T_p$ corresponds to the largest oscillation period of the system, and numerically compute the average values of these quantities.

\begin{figure}
	\includegraphics[width = \linewidth]{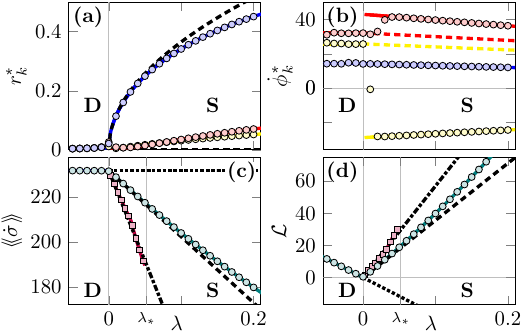}
		\caption{Comparison of analytic results (dashed lines), macroscopic dynamics (solid lines), and numerical simulations (symbols, $N=10^6$) for $q=\beta f =7$ in decoherent phase (D) and synchronised phase (S), averaged over long times, $T\approx 100\tau$. (a) Order parameters $r^*_1$, $r^*_2$, and $r^*_3$ in yellow, red, and blue, respectively, for Arrhenius rates. (b) Oscillation frequencies $\dot\phi^*_1$, $\dot\phi^*_2$, and $\dot\phi^*_3$ in yellow, red, and blue, respectively, for Arrhenius rates. (c) Average entropy production rate as function of $\lambda$ for Arrhenius (bullets) and Glauber (squares) rates. The dotted line shows $\langle\dot\sigma\rangle_0$, the dashed and dash-dotted lines show the asymptotic behaviours as obtained from \Eqnref{dO}. (d) Average phase-space contraction rate as function of $\lambda$ for Arrhenius (bullets) and Glauber (squares) rates. The dotted line shows $\mc{L}_0$, the dashed and dash-dotted lines show the asymptotic behaviours as obtained from \Eqnref{L}.}\figlab{numsim_arrh}
\end{figure}

Figure~\figref{numsim_arrh} shows the results of our numerical simulations (symbols) for $q=7$, compared with the theory, shown as the lines. Figure~\figref{numsim_arrh}(a) shows the amplitudes $\ve r^*$ in the decoherent state (D, $\lambda<0$) and in the synchronised state (S, $\lambda>0$) for Arrhenius rates, for which $S_a = \{3\}$, according to the theory developed in Sec.~\secref{amp}. We observe that only $r^*_3$ changes significantly across the transition into the synchronised state; all other amplitudes remain small. The analytical predictions (dashed lines), agree well with both the numerics and with the results from the macroscopic dynamics (solid lines) close to the bifurcation.

Figure~\figref{numsim_arrh}(b) shows the components of the angular velocity $\ve{\dot \phi}$ as function of $\lambda$ for Arrhenius rates. In the decoherent phase, the different modes oscillate with their natural frequencies $\ve \omega$, but the amplitudes of these oscillations are essentially zero, cf. \Figref{numsim_arrh}(a). In the synchronised state, the angular velocity $\dot\phi_3$ of the active mode $z_3$ is slightly reduced, according to \Eqnref{relphase}, while all other velocities are altered significantly. Notably, $\dot\phi_1$ changes sign, as we already noted for $q=9$ in \Figref{oscillations}(j). Close to the synchronisation transition, we observe that for $\dot\phi_1$ and $\dot\phi_2$ the angular velocities result from a mixture of the velocities of the (unstable) modes $z_1$ and $z_2$ (dashed lines) and those of the actual velocities of the active mode $z_3$ (solid lines). This is a finite size effect that vanishes in the thermodynamic limit.

Figure~\figref{numsim_arrh}(c) shows the time-averaged dissipation $\llangle \dot \sigma\rrangle$ in the decoherent and synchronised states from numerical simulations using Arrhenius (bullets) and Glauber (squares) rates, and from theory (lines). As predicted, dissipation is reduced by synchronisation, compared to the decoherent state (dotted line) that becomes unstable for $\lambda>0$. The analytical predictions for $\llangle \dot \sigma\rrangle$ close to the bifurcation (dashed and dash-dotted lines) are in agreement with the numerical simulations. For larger $\lambda$, we observe deviations from the predicted behaviour. In particular, for Glauber rates at $\lambda_*\approx0.05$ (light grey line), the system undergoes an additional bifurcation into an ordered state, where dissipation is drastically reduced.

Figure~\figref{numsim_arrh}(d) shows the phase-space contraction rate $\mc{L}$ from numerical simulations using Arrhenius (bullets) and Glauber (squares) rates, in comparison with the analytical predictions (dashed and dash-dotted lines) and the result using the macroscopic dynamics (solid lines). We observe that $\mc{L}$ vanishes at the phase transition, $\lambda=\Lambda=0$, but is otherwise positive. For $\lambda>0$, the decoherent fixed point is unstable and its phase-space contraction $\mc{L}_0$ is negative (dotted line). In the synchronised phase for $\lambda>0$, we observe positive $\mc{L}$, in good agreement between the analytical prediction \eqnref{dL} (dashed and dash-dotted lines), numerical simulations (bullets and squares), the results from the macroscopic dynamics (solid lines). Larger $\lambda$ result in deviations from the analytical results, obtained from the normal form. For Glauber rates, the bifurcation to the ordered state at $\lambda_*$ leads to a drastic increase in stability. The data presented in \Figsref{numsim_arrh}(c) and (d) have been used to test the stability-dissipation relation~\eqnref{sdr} (see Fig.~4 in Ref.~\cite{Mei24a}).

Multistability, predicted for Arrhenius rates in \Figref{phasediag_2}, is numerically confirmed for $q=9$ in \Figref{Arrhenius_lc}.
\begin{figure}[t]
	\includegraphics[width=\linewidth]{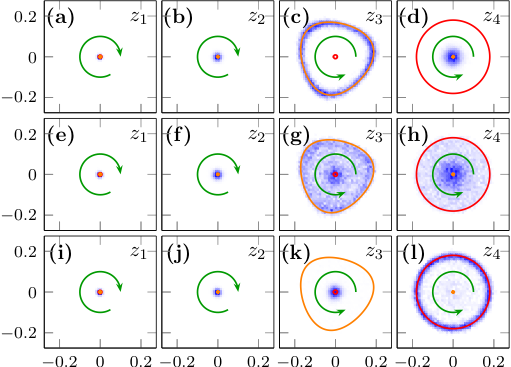}
	\caption{Multistability for Arrhenius dynamics. Probability density (blue) of $z_1,\ldots,z_4$ in the complex plane, for $q=9$ with $\lambda = 0.025$, $\beta f = 7$, $N=10^5$, sampled from an ensemble of $10^4$ realisations initialised close to the synchronised state with $S_a = \{3\}$. The lines show the prediction from the macroscopic dynamics~\eqnref{eom} $S_a=\{3\}$ state (orange) and the more stable $S_a=\{4\}$ state (red). Arrows show the directions of motion. (a)--(d): Initial state. (e)--(h): State after $t=5\tau$. (i)--(l): State after $t=15\tau$.}\figlab{Arrhenius_lc}
\end{figure}
We initialise an ensemble of realisations of the system close to the stable synchronised state with $S_a=\{3\}$ [\Figsref{Arrhenius_lc}(a)--\figref{Arrhenius_lc}(d)]. Initially, the density remains at the $S_a=\{3\}$ state for several relaxation times $\tau$. As time evolves [\Figsref{Arrhenius_lc}(e)--\figref{Arrhenius_lc}(h)], however, an increasing fraction of the density transitions into the more stable (and less dissipative) synchronised state with $S_a=\{4\}$. Eventually [\Figsref{Arrhenius_lc}(i)--\figref{Arrhenius_lc}(l)] the majority of the realisations occupy the $S_a=\{4\}$ state. We have conducted a similar analysis for the same system but with $q=17$, where three synchronised states ($S_a=\{6\}$, \{7\}, and \{8\}) are stable. The results can found be in the Supplemental Material of Ref.~\cite{Mei24a}. They confirm that finite systems close to the transition eventually predominantly occupy the least dissipating small-amplitude state, independently of the initial conditions, in accordance with the minimum-dissipation principle discussed in \Secref{sssel} and in Ref.~\cite{Mei24a}.

From these numerical results, it appears that the minimum-dissipation principle, the selection of states based on minimum entropy production, holds for much larger systems than allowed by the condition $1\ll N\lesssim \lambda^{-2}$ derived in \Secref{sssel}. This indicates that in presence of multistability, (minimum) entropy production plays a role for the transition rates between different attractors, even when these transitions are rare.

Finally, for Glauber dynamics ($a=1$), the phase diagram in~\Figref{phasediag} predicts stable coherent oscillations with an increasing number of active modes for increasing $q$. In particular for $q=11$, theory claims the existence of a unique stable synchronised state with two active Fourier modes $S_a = \{4,5\}$ and $|S_a|=2$. This finding is confirmed by our numerical simulations shown in \Figref{Glauber_lc}. We clearly observe two active Fourier modes, the ones with largest $k$, as predicted. The amplitudes of the other modes are small.

To summarise, our numerical simulations confirm the predictions of the theory close to the synchronisation transition. Further away from the transition, we observe deviations from the predicted behaviour and additional bifurcations, indicating a more complex behaviour than accounted for by the normal form~\eqnref{bhnf}, when $\Lambda$ (and thus $\lambda$) is not small.
\section{Conclusions}\seclab{conc}
\begin{figure}
	\includegraphics[width=\linewidth]{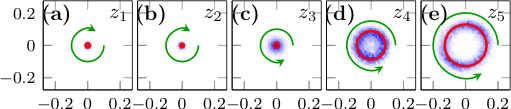}
	\caption{Multiple active modes for Glauber dynamics. Probability density (blue) of $z_1,\ldots,z_5$ in the complex plane, for $q=11$ with $\lambda = 0.01$, $\beta f = 7$, $N=10^5$, sampled from an ensemble of $10^4$ realisations at $t=4\tau$, initialised from the decoherent state. The active modes are $S_a = \{4,5\}$. The red line shows the prediction from the macroscopic dynamics. The arrows show the directions of motion.}\figlab{Glauber_lc}
\end{figure}

We have studied small-amplitude synchronisation and stationary states in driven Potts models with general dynamics and all numbers $q$ of states. These models describe ferromagnetically interacting Potts spins in contact with a heat bath at inverse temperature $\beta$ and driven by a non-conservative force $f$. The competition between decoherence from thermal noise and alignment due to ferromagnetic interactions, combined with non-equilibrium driving, results in a dynamical phase transition from the decoherent phase into either a synchronised phase or, in some cases, non-equilibrium stationary states.

To study the transition, we have derived and analysed the normal form of the high-dimensional Hopf bifurcation that underlies it. Close to the transition, we found stable synchronised and stationary states of small amplitude for a wide range of parameters. The stability of the decoherent phase as well as the characteristics of the small-amplitude states were found to depend on the dynamics in an intricate way.

In particular, dynamics may stabilise or destabilise the decoherent phase, compared to the equilibrium Potts model, depending on the location in parameter space where the bifurcation parameter $\Lambda$ in \Eqnref{lamom} changes sign. In the synchronised phase, $q$ and $a$ [cf. \Eqnref{aparam}] span the parameter space of a rich phase diagram (\Figref{phasediag}) of coherent small-amplitude oscillations.

Each synchronised state hosts a finite number of active Fourier modes that determine the shape of the probability distribution of oscillator states in the thermodynamic limit. The number of active modes was found to be much smaller than the maximum possible, with active modes being the ones with largest $k$. This shows that driven Potts oscillators synchronise differently than oscillators in the Kuramoto model: While the phases in Kuramoto oscillators localise in real space, Potts oscillators localise in Fourier space.

Connecting to the thermodynamics of the model, we found that in driven Potts models synchronisation reduces dissipation, independently of the dynamics and of the number $q$ of states within each oscillator. Furthermore, we have shown that dissipation is minimised in the most stable synchronised state. Close to the synchronisation transition, this finding is summarised in a linear stability-dissipation relation, given \Eqnref{sdr} and discussed in Ref.~\cite{Mei24a}. At finite system size and close to the transition, $1\ll N\lesssim\lambda^{-2}$, our findings suggest the existence of a minimum-dissipation principle for driven Potts models, where the system probabilistically selects the least dissipating state based on maximum stability, i.e., maximum phase-space contraction. In practice, our numerical results indicate that the minimum-dissipation principle holds for significantly larger $N$ and $\lambda$ than explained by this estimate.

Our analysis opens the door to a deeper understanding of driven Potts models, synchronisation transitions, and non-equilibrium phase transitions in general. The phase diagram in~\Figref{phasediag} is an ideal starting point to study regimes outside the vicinity of the transition. We have conducted a preliminary study that indeed shows that the model exhibits additional transitions into quasiperiodic and possibly chaotic, synchronised states for large-enough $\lambda$. It would be interesting to investigate systematically how these additional transitions affect the thermodynamics~\cite{Gas20}, and, in particular, whether or not they reduce dissipation even further.

On the basis of the deterministic coherent oscillations we have studied here, another logical next step is the analysis of finite-$N$ fluctuations. This would allow us to directly quantify the coherence of oscillations~\cite{Gas02a,Gas02b,Bar17}, and to estimate the rates of transitions between stable small-amplitude states. This could provide a quantitative estimate of the probabilities to reside in given stable states, and thus possibly extend the minimum-dissipation principle to regimes outside the vicinity of the phase transition.

The linear stability-dissipation relation~\eqnref{sdr} has been shown to hold for all driven Potts models and all $q$, and might, with possibly model-specific adjustments, hold even for more general classes of models. Intriguing candidates to test this hypothesis are driven lattice Potts models with nearest-neighbour interactions~\cite{Woo06}, for which preliminary studies show promising results, and perhaps also active versions of the model~\cite{Man23}. Driven lattice Potts models close to the synchronisation transition could exhibit interesting critical fluctuations, whose characteristics might affect the minimum-dissipation principle and the stability-dissipation relation established here.
\begin{acknowledgments}
	We thank Gianmaria Falasco for insightful discussions and for pointing out the notion of phase-space contraction rate in stochastic systems.
	JM's stay at King's College London was supported by a Feodor-Lynen scholarship of the Alexander von Humboldt-Foundation. 
        ME was supported by the ChemComplex project (C21/MS/16356329) funded by FNR (Luxembourg).
\end{acknowledgments}

\appendix
\section{Coordinate transform}\seclab{appctrafo}
We consider the effect of a transformation $\ve{\hat x}\to\ve{\hat x'}$ that removes the quadratic terms $\ve{\hat h}^{(2)}$ in \Eqnref{eomexp}. We write
\algn{\eqnlab{xtrans}
	\ve{\hat x'} \equiv \ve{F}(\ve{\hat x}) = \ve{\hat x} + \ve f^{(2)}(\ve{\hat x})\,.
}
The inverse transformation reads
\algn{\eqnlab{Finv}
	\ve{\hat x} = \ve{F}^{-1}(\ve{\hat x'}) \sim \ve{\hat x'} - \ve f^{(2)}(\ve{\hat x'}) +[\ve f^{(2)}(\ve{\hat x'})\cdot\nabla]\ve f^{(2)}(\ve{\hat x'})\,.
}
To obtain the flow of $\ve{\hat x'}$, we take a time derivative of \Eqnref{xtrans}, and write
\algn{
	\dd{t}\ve{\hat x'} 	&= [\ve{\hat h}(\ve{\hat x}) \cdot\nabla]\ve F(\ve{\hat x})\,,\nn\\
					&= \{\ve{\hat h}[\ve{F}^{-1}(\ve{\hat x'})] \cdot\nabla\}\ve F[\ve{F}^{-1}(\ve{\hat x'})]\,. \eqnlab{dxtilde}
}
The term $\ve{\hat h}[\ve{F}^{-1}(\ve{\hat x'})]$ reads to third order in $\ve{\hat x'}$,
\begin{multline}
	\ve{\hat h}[\ve{F}^{-1}(\ve{\hat x'})]\sim \ve{\hat h}^{(1)} - (\ve f^{(2)}\cdot\nabla) \ve{\hat h}^{(1)}+  [(\ve f^{(2)}\cdot\nabla)\ve f^{(2)}\cdot\nabla]\ve{\hat h}^{(1)}\\
	+ \ve{\hat h}^{(2)} - (\ve f^{(2)}\cdot\nabla) \ve{\hat h}^{(2)} + \ve{\hat h}^{(3)}\,,
\end{multline}
where all functions have $\ve{\hat x'}$ as their argument. We use the notation $(D\ve{G})_{kk'} = \partial G_k/\partial \hat x'_{k'}$ for a vector $\ve G$ and write
\algn{\eqnlab{DF}
	D\ve F(\ve{\hat x}) \sim \ve{I} + D\ve f^{(2)}(\ve{\hat x}) + D\ve f^{(3)}(\ve{\hat x})\,.
}
Here, $\boldsymbol I$ denotes the $q\times q$ identity matrix. We now express \Eqnref{dxtilde} as
\algn{\eqnlab{xheom}
\dd{t}\ve{\hat x'}		&= \ve{\hat h}[\ve{F}^{-1}(\ve{\hat x'})]^{\sf T}D\ve F(\ve{F}^{-1}(\ve{\hat x'}))\,,
}
where the transpose indicates that the vector $\ve{\hat h}$ multiplies the matrix $D\ve F$ from the left.

Using \Eqsref{Finv} and \eqnref{DF} we express $D\ve F[\ve{F}^{-1}(\ve{\hat x'})]$ as
\algn{
	D\ve F(\ve{F}^{-1}(\ve{\hat x'})) 	\sim \ve{I} + D\ve f^{(2)} - (\ve f^{(2)}\cdot\nabla) D\ve f^{(2)}
}
Keeping only terms up to third order in $\ve{\hat x'}$, we write \Eqnref{xheom} as
\begin{multline}\eqnlab{xhexp}
	\dd{t}\ve{\hat x'}	 \sim \ve{\hat h}^{(1)} - (\ve f^{(2)}\cdot\nabla) \ve{\hat h}^{(1)}  + \ve{\hat h}^{(1)\sf T} D\ve f^{(2)} + \ve{\hat h}^{(2)}\\
	+\ve{\hat h}^{(2)\sf T}D\ve f^{(2)}- \ve{\hat h}^{(1)\sf T}(\ve f^{(2)}\cdot\nabla) D\ve f^{(2)} - (\ve f^{(2)}\cdot\nabla) \ve{\hat h}^{(1)\sf T}D\ve f^{(2)} \\
	+ [(\ve f^{(2)}\cdot\nabla)\ve f^{(2)}\cdot\nabla]\ve{\hat h}^{(1)} - (\ve f^{(2)}\cdot\nabla) \ve{\hat h}^{(2)} + \ve{\hat h}^{(3)}\,.
\end{multline}
To remove the quadratic terms in the first line of \Eqnref{xhexp}, we require that~\cite{Guc83,Cra91a,Cra91b}
\algn{\eqnlab{L2cond}
	\ve{\hat h}^{(2)} = (\ve f^{(2)}\cdot\nabla) \ve{\hat h}^{(1)}  - \ve{\hat h}^{(1)\sf T} D\ve f^{(2)} \equiv \mbb{L}(\ve f^{(2)})\,.
}
This is \Eqnref{L2eqn} in the main text. Applying $\ve{f}^{(2)}\cdot\nabla$ to \Eqnref{L2cond} we furthermore find the relation~\cite{Guc83,Cra91a,Cra91b}
\begin{multline}
	 (\ve f^{(2)}\cdot\nabla) \ve{\hat h}^{(2)}=[(\ve f^{(2)}\cdot\nabla)\ve f^{(2)}\cdot\nabla]\ve{\hat h}^{(1)}\\
	  - (\ve f^{(2)}\cdot\nabla) \ve{\hat h}^{(1)\sf T}D\ve f^{(2)} - \ve{\hat h}^{(1)\sf T}(\ve f^{(2)}\cdot\nabla) D\ve f^{(2)}\,,
\end{multline}
which then simplifies \Eqnref{xheom} to
\algn{
	\dd{t}\ve{\hat x'}	 = \ve{\hat h}^{(1)}+\ve{\hat h}^{(2)\sf T}D\ve f^{(2)} + \ve{\hat h}^{(3)}\,.
}
This means that a transformation that removes $\ve{\hat h}^{(2)}$ by satisfying \Eqnref{L2cond} changes the third-order terms of the flow of $\ve{\hat x}$ by
\algn{
	\ve{\hat g}^{(3)} = \ve{\hat h}^{(3)} + \ve{\hat h}^{(2)\sf T}D\ve f^{(2)}\,,
}
as stated in \Eqnref{g3eqn} in the main text.
\section{General derivation of normal-form equations}\seclab{gender}
Here we derive the normal form given in \Eqnref{bhnf} in the main text. The discrete Fourier (back) transforms of $\ve{\hat x}$ and $\ve{x}$ are given by
\algn{
	\hat x_k  = \sum_{n=0}^{q-1}\ee^{\frac{i2\pi k n}{q}}x_n\,,\quad 	x_n  = \frac1q\sum_{k=0}^{q-1}\ee^{-\frac{i2\pi k n}{q}}\hat x_k
}
The equation of motion~\eqnref{eom} for $\ve x$ reads
\algn{\eqnlab{xnflow}
	\dot x_n = j(x_{n},x_{n-1}) - j(x_{n+1},x_{n})\,.
}
Expanding the average flux $j$ to third order in $\ve x$ we find
\algn{
	j(y,x) =& j_{00} + j_{10}y + j_{01}x + \frac12\left(j_{20}y^2 + 2j_{11}xy + j_{02}x^2\right)\nn\\
		&+\frac16\left(j_{30}y^3 + 3j_{21}xy^2 + 3j_{12}x^2 y + j_{03}x^3\right)\,.
}
We note that
\algn{
	j(x,x) = (k^+-k^-)x\,,
}
is linear in $x$.  This implies the relations
\algn{
	j_{00} =& \frac{k^+-k^-}q\,,\quad j_{10} + j_{01} = q j_{00}\,,\nn\\
	j_{20} + j_{02} =& -2j_{11}\,,\quad  j_{30} + j_{03} = -3\left(j_{21} + j_{12}\right)\,.\eqnlab{jrels}
}
The general form~\eqnref{flux} of the fluxes also implies that
\algn{\eqnlab{jrels2}
	j_{03}-j_{30} = j_{21}-j_{12}\,,
}
which can be checked by explicit computation. The flow~\eqnref{xnflow} then reads to third order in $\ve x$,
\algn{
	&\dot x_n \sim \left(j_{10}-j_{01}\right)x_n+j_{01}x_{n-1}-j_{10}x_{n+1}\nn\\
		&+\frac12\left(j_{20}x_n^2 + 2j_{11}x_nx_{n-1} + j_{02}x_{n-1}^2\right)\nn\\
		&-\frac12\left(j_{20}x_{n+1}^2 + 2j_{11}x_{n+1}x_n + j_{02}x_n^2\right)\nn\\
		&+\frac16\left(j_{30}x_n^3 + 3j_{21}x_{n}^2x_{n-1} + 3j_{12}x_{n}x_{n-1}^2+ j_{03}x_{n-1}^3\right)\nn\\
		&-\frac16\left(j_{30}x_{n+1}^3 + 3j_{21}x_{n+1}^2x_n + 3j_{12}x_{n+1}x_n^2 + j_{03}x_n^3\right)\,.
}
We now Fourier transform this expression to obtain the flow of $\ve{\hat x}$. For the linear terms we find
\algn{
	\dot{\hat x}_k 	&= \left(\mu_k + i\omega_k	\right)\hat x_k\,.
}
The quadratic terms in $\dot x_n$ transform according to
\sbeqs{\eqnlab{x2ftrans}
\algn{
	\sum_{n=0}^{q-1}\ee^{\frac{i2\pi k n}{q}}x^2_n =& \frac1{q}\sum_{k'=0}^{q-1}{\hat x}_{k'}{\hat x}_{k-k'}\,,\\
	\sum_{n=0}^{q-1}\ee^{\frac{i2\pi k n}{q}}x^2_{n\pm1} =& \frac{\ee^{\mp\frac{ i2\pi k}{q}}}{q}\sum_{k'=0}^{q-1}{\hat x}_{k'}{\hat x}_{k-k'}\,,\\
	\sum_{n=0}^{q-1}\ee^{\frac{i2\pi k n}{q}}x_{n\pm1}x_n =& \frac1{q} \sum_{k'=0}^{q-1}\ee^{\mp\frac{i2\pi k'}{q}}{\hat x}_{k'}{\hat x}_{k-k'}
}
}
These quadratic contributions to the flow of $\hat x_k$ are then
\algn{\eqnlab{hhat2}
	{\hat h}^{(2)}_k(\ve{\hat x})	= 	\frac1{q} \sum_{k'=1}^{q-1}\left(\frac{j_{20}-j_{02}}{\Lambda}\mu_{k} - i\frac{j_{11}}{qj_{00}}\omega_{kk'}\right){\hat x}_{k'}{\hat x}_{k-k'}\,,
}
where we defined $\omega_{kk'} \equiv \omega_{k} -\omega_{k'}-\omega_{k-k'}$.

The third-order terms transform as
\algn{
	\sum_{n=0}^{q-1}\ee^{\frac{i2\pi k n}{q}}x^3_n =& \frac1{q^2}\sum_{k',k''=1}^{q-1}{\hat x}_{k'}{\hat x}_{k''}{\hat x}_{k-k'-k''}\,,\nn\\
	\sum_{n=0}^{q-1}\ee^{\frac{i2\pi k n}{q}}x^3_{n\pm1} =& \frac{\ee^{\mp\frac{ i2\pi k}{q}}}{q^2}\sum_{k',k''=1}^{q-1}{\hat x}_{k'}{\hat x}_{k''}{\hat x}_{k-k'-k''}\,,\nn\\
	\sum_{n=0}^{q-1}\ee^{\frac{i2\pi k n}{q}}x^2_n x_{n\pm1} =& \frac1{q^2} \sum_{k',k''=1}^{q-1}\ee^{\mp\frac{i2\pi k'}{q}}{\hat x}_{k'}{\hat x}_{k''}{\hat x}_{k-k'-k''}\,,\nn\\
	\sum_{n=0}^{q-1}\ee^{\frac{i2\pi k n}{q}}x_n x^2_{n\pm1} =& \frac1{q^2} \sum_{k',k''=1}^{q-1}\ee^{\mp\frac{i2\pi (k'+k'')}{q}}{\hat x}_{k'}{\hat x}_{k''}{\hat x}_{k-k'-k''}\,.
}
This gives
\algn{
	&{\hat h}^{(3)}_k	= \frac1{6q^2}\sum_{k',k''=1}^{q-1}\left[2\frac{j_{03}-j_{30}}{\Lambda}\left(3\mu_{k-k'}-3\mu_{k'}-\mu_k\right)\right.\nn\\
	&\left.+ i\frac{j_{30}+j_{03}}{qj_{00}}\omega_{kk'}\right]{\hat x}_{k'}{\hat x}_{k''}{\hat x}_{k-k'-k''}\,.\eqnlab{hhat3}
}
In \Eqnref{hhat2}, we note that $\ve{e}_{k'}$ given in \Eqnref{basel} in the main text are equivariant basis vectors, with respect to which $\mbb{L}$ is diagonal. The transformation that removes the quadratic terms in ${\hat h}^{(2)}_k(\ve{\hat x})$ is given by
\algn{
	{\hat f}^{(2)}_k(\ve{\hat x})	= -\frac1{q} \sum_{k'=1}^{q-1}\left(\frac{j_{11}}{qj_{00}} + i\frac{j_{20}-j_{02}}{\Lambda}\frac{\mu_k}{\omega_{kk'}}\right){\hat x}_{k'}{\hat x}_{k-k'}\,,
}
expanded to lowest order in $0<\Lambda\ll1$ and $\Lambda\ll\beta f$.

The cubic contribution generated by ${\hat f}^{(2)}_k$ reads
\algn{
	(\ve{\hat h}^{(2)\sf T}D\ve f^{(2)})_k = \sum_{k'=1}^{q-1}\hat h^{(2)}_{k'}\partial_{k'} {\hat f}^{(2)}_k\,,
}
where
\algn{
	\partial_{k'} {\hat f}^{(2)}_k \sim -\frac2{q} \left[\frac{j_{11}}{qj_{00}} + i\frac{j_{20}-j_{02}}{\Lambda}\frac{\mu_k}{\omega_{kk'}}\right]{\hat x'}_{k-k'}\,.
}
We thus find
\algn{
	&(\ve{\hat h}^{(2)\sf T}D\ve f^{(2)})_k	=-\frac1{q^2}\sum_{k',k''=1}^{q-1}\left\{\frac{2(j_{20}-j_{02})j_{11}}{qj_{00}\Lambda}\left(\mu_{k'}+\frac{\mu_k\omega_{k'k''}}{\omega_{kk'}}\right)\right.\nn\\
				&+ \left.\frac{i}2\left[\left(2\frac{j_{20}-j_{02}}{\Lambda}\right)^2\frac{\mu_k\mu_{k'}}{\omega_{kk'}} - \left(\frac{2j_{11}}{qj_{00}}\right)^2\omega_{k'k''}\right]\right\}{\hat x'}_{k''}{\hat x'}_{k'-k''}{\hat x'}_{k-k'}\,. \eqnlab{ghat3}
}
In the next step, we need to extract the parts of $\ve{\hat h}^{(2)\sf T}D\ve f^{(2)}$ and $\ve{\hat h}^{(3)}$ that are of the form $\hat x_k \hat x_{k'}\hat x_{-k'}$, because these are the essential terms for the bifurcation. To combine these terms in a single sum, we need to subtract the terms that are contained in both sums in \Eqnref{ghat3}. For these terms, $k''=k$ and $k'=2k$. We then transform $k'\to k-k'$, which means that we must remove terms with $k=k'$. This gives for $k\neq\frac{q}2$,
\begin{widetext}
\algn{
	&(\ve{\hat h}^{(2)\sf T}D\ve f^{(2)})_k 	\propto -\frac1{q^2}\sum_{k'=1}^{\lfloor\frac{q-1}2\rfloor}\left(1-\frac12\delta_{kk'}\right)\left\{\frac{4(j_{20}-j_{02})j_{11}}{qj_{00}\Lambda}\tilde\mu_{kk'}+ i\left[\left(\frac{j_{20}-j_{02}}{qj_{00}}\right)^2\omega_k + \left(\frac{2j_{11}}{qj_{00}}\right)^2\tilde\omega_{kk'} \right]\right\}{|z'_{k'}|^2}{z'}_{k}\nn\\
	&-\frac1{2q^2}\frac{4(j_{20}-j_{02})j_{11}}{qj_{00}\Lambda}\mu_{k}{|z'_{k}|^2}{z'}_{k}-\frac1{2q^2}\left\{\frac{4(j_{20}-j_{02})j_{11}}{qj_{00}\Lambda}\tilde\mu_{k\frac{q}2}+ i\left[\left(\frac{j_{20}-j_{02}}{qj_{00}}\right)^2\omega_k + \left(\frac{2j_{11}}{qj_{00}}\right)^2\tilde\omega_{k\frac{q}2}\right]\right\}{z'}^3_{\frac{q}2}\,,\eqnlab{g3}
}
\end{widetext}
where we defined $\tilde \mu_{kk'} \equiv \mu_{k-k'} + \mu_{k+k'}-2\mu_k$ and $\tilde \omega_{kk'} \equiv 2\omega_{k} - \omega_{k-k'}-\omega_{k+k'}$.

We now apply the same procedure to $\ve{\hat h}^{(3)}$. The terms $\propto\hat x_k \hat x_{k'}\hat x_{-k'}$ in the sum in \Eqnref{hhat3} come from terms with $k'=k$, $k''=k$ and $k'+k'' = 0$, so we need to subtract the overlap terms with $k'=k''=k$, $k'=-k''=k$, and $-k'=k''=k$. We have
\begin{widetext}
\algn{
	{\hat h}^{(3)}_k	&\propto \frac1{q^2}\sum_{k'=1}^{\lfloor \frac{q-1}2\rfloor}\left(1-\frac12\delta_{kk'}\right)\left\{2\frac{j_{03}-j_{30}}{\Lambda}\left(\tilde\mu_{kk'}-2\mu_{k'}\right) + \frac{i}3\frac{(j_{30}+j_{03})}{qj_{00}}\tilde\omega_{kk'}\right\}|z'_{k'}|^2z'_{k}\nn\\
	&+\frac1{2q^2}\left\{2\frac{j_{03}-j_{30}}{\Lambda}\left(\tilde\mu_{k\frac{q}3}-2\mu_{\frac{q}2}\right) + \frac{i}3\frac{j_{30}+j_{03}}{qj_{00}}\tilde\omega_{k\frac{q}2}\right\}{z'}^3_{\frac{q}2}\,.\eqnlab{h3}
	}
\subsection{Normal-form equations}\seclab{appbhnf}
Putting expressions~\eqnref{g3} and \eqnref{h3} together and rearranging the terms, we find
\algn{
&{\hat g}^{(3)}_k	\propto \frac1{q^2}\sum_{k'=1}^{\lfloor \frac{q-1}2\rfloor}\left\{-\left[1-\frac12\left(1-\frac{j_{20}-j_{02}}{j_{03}-j_{30}}\frac{j_{11}}{qj_{00}}\right)\delta_{kk'}\right]4\frac{j_{03}-j_{30}}{\Lambda}\mu_{k'}+\left(1-\frac12\delta_{kk'}\right)\left[2\left(\frac{j_{03}-j_{30}}{\Lambda}-\frac{8(j_{20}-j_{02})j_{11}}{qj_{00}\Lambda}\right)\tilde\mu_{kk'}\right.\right.\nn\\
	&\left.\left.-i\left(\frac{j_{20}-j_{02}}{qj_{00}}\right)^2\omega_k+ i\left[\frac13\frac{j_{30}+j_{03}}{qj_{00}}-\left(\frac{2j_{11}}{qj_{00}}\right)^2\right]\tilde\omega_{kk'}\right]\right\}|z'_{k'}|^2z'_{k}+\frac1{2q^2}\left\{ -4\frac{j_{03}-j_{30}}{\Lambda}\mu_{\frac{q}2}+ 2\left(\frac{j_{03}-j_{30}}{\Lambda}-\frac{8(j_{20}-j_{02})j_{11}}{qj_{00}\Lambda}\right)\tilde\mu_{k\frac{q}2}\right.\nn\\
	&\left.-i\left(\frac{j_{20}-j_{02}}{qj_{00}}\right)^2\omega_k+ i\left[\frac13\frac{j_{30}+j_{03}}{qj_{00}}-\left(\frac{2j_{11}}{qj_{00}}\right)^2\right]\tilde\omega_{k\frac{q}2}\right\}{z'}^3_{\frac{q}2}\,.
}
For $k=\frac{q}2$ the expressions are slightly different; we find
\algn{
{\hat g}^{(3)}_{\frac{q}2}&\propto\frac1{q^2}\sum_{k'=1}^{\lfloor \frac{q-1}2\rfloor}\left\{-8\left(\frac{j_{03}-j_{30}}{\Lambda}-\frac{(j_{20}-j_{02})j_{11}}{qj_{00}\Lambda}\right)\mu_{k'}\right\}|z'_{k'}|^2{z'}_{\frac{q}2}-\frac{4}{3q^2}\frac{j_{03}-j_{30}}{\Lambda}\mu_{\frac{q}2}{z'}^3_{\frac{q}2}\,.
}
In summary,
\algn{\eqnlab{h3g3k}
{\hat g}^{(3)}_k	&\propto -\sum_{k'=1}^{\lfloor \frac{q-1}2\rfloor}\left\{A\left[1+\frac14\left(	1-\frac{C}A\right)\delta_{kk'}\right]\sin^2\left(\frac{\pi k'}{q}\right)+\left(1-\frac12\delta_{kk'}\right)\left[C\cos\left(\frac{2\pi k}{q}\right)\sin^2\left(\frac{\pi k'}{q}\right)\right.\right.\nn\\
	&\left.\left.+iB\sin\left(\frac{2\pi k}{q}\right)+ iD\sin\left(\frac{2\pi k}{q}\right)\sin^2\left(\frac{\pi k'}{q}\right)\right]\right\}|z'_{k'}|^2z'_{k}-\frac1{2}\left\{A  + C\cos\left(\frac{2\pi k}{q}\right) + i(B+D)\sin\left(\frac{2\pi k}{q}\right)\right\}{z'}^2_{\frac{q}2}z'_k
}
\end{widetext}
for $k\neq\frac{q}2$ and
\algn{\eqnlab{h3g3q2}
{\hat g}^{(3)}_{\frac{q}2}&\propto-\sum_{k'=1}^{\lfloor \frac{q-1}2\rfloor}(A-C)\sin^2\left(\frac{\pi k'}{q}\right)|z'_{k'}|^2z'_{k}-\frac{A}{3}{z'}^3_{\frac{q}2}
}
for $k = \frac{q}2$, with coefficients
\algn{
	A &= \frac{4}{q^2}(j_{03}-j_{30})\,,\quad C = \frac{4}{q^2}\left[\frac{(j_{02}^2-j_{20}^2)}{qj_{00}}-(j_{03}-j_{30})\right]\,,\nn\\
	B &= \frac1{q^2}\frac{(j_{20}-j_{02})^2}{qj_{00}}\,,\quad D = \frac{4}{q^2}\left[\frac{(j_{20}+j_{02})^2}{qj_{00}}-\frac13(j_{30}+j_{03})\right]\,.
}
In the final step, we combine the terms in the sums in \Eqsref{h3g3k} and \eqnref{h3g3q2} into the matrix $\mbb{C}_{kk'}$ and define $a\equiv C/A$ and $b\equiv D/B$, to obtain the normal form in~\Eqnref{bhnf} in the main text.
\section{Stable oscillations for $a<\frac13$}\seclab{appamp}
In this Section, we show that for $a<\frac13$ the stability matrix $\mbb{N}$ has at least $|S_a|-1$ positive eigenvalues, as a consequence of the matrix determinant lemma~\cite{Bro20} and an eigenvalue interlacing theorem~\cite{Bun78,Tho76,Wil65}.

As explained in the main text, we can bring the stability matrix $\mbb{N}$ in \Eqnref{stabmat} into a form where the elements $l,m\in S_a$ form a connected block $\mbb{N}_a$, and $\mbb{N}_{ia}$ gives the diagonal part. The eigenvalues of $\mbb{N}$ are then given by the diagonal elements in $\mbb{N}_{ia}$ and the eigenvalues of the block $\mbb{N}_{a}$.

We now show that the connected bloc $\mbb{N}_{a}$ has $|S_a|-1$ positive eigenvalues for $a<\frac13$. To this end, we symmetrise $\mbb{N}$, writing $\mbb{\tilde N} = \mbb{S}\mbb{N}\mbb{S}^{-1}$, where $\mbb{S}$ is diagonal with elements
\algn{
	\mbb{S}_{kk'} = \sqrt{\frac{\tilde v_k}{\tilde u_{k}}}\delta_{kk'}\,.
}
This transformation does not affect the eigenvalues and leaves the diagonal part of $\mbb{N}$ invariant. The block $\mbb{N}_a$, however, transforms into the sum of a diagonal part and a symmetric order-one perturbation,
\algn{\eqnlab{rankonepert}
	\mbb{\tilde N}_{a} = -(\ve{\tilde w}_a\ve{\tilde w}_a^{\sf T}-\mbb{\tilde W}_a)\,,
}
with the symmetric rank-one perturbation $\ve{\tilde w}_a\ve{\tilde w}_a^{\sf T}$ characterised by the vector $\tilde w$ with components
\algn{
	\tilde w_k = \sqrt{\tilde u_k \tilde v_k}\,.
}
For matrices of the form~\eqnref{rankonepert}, the Matrix determinant lemma~\cite{Bro20} states that the determinant of $\mbb{\tilde N}_a$, and thus of $\mbb{N}_a$, is given by
\algn{\eqnlab{Kdet}
	\det\mbb{\tilde N}_a &= \det(\mbb{\tilde W}_a)\left(1-\ve{\tilde w}_a^{\sf T}\mbb{\tilde W}_a^{-1}\ve{\tilde w}_a \right)
}
The elements of $\mbb{\tilde W}_a$ are positive whenever all elements of $\mbb{W}_a$ are positive, which is guaranteed for $a<\frac13$. A simple analysis shows that the term $1-\ve{\tilde w}_a^{\sf T}\mbb{\tilde W}_a^{-1}\ve{\tilde w}_a$ is negative for $a<\frac13$, which implies that the determinant of $\mbb{\tilde N}_a$ is negative. Hence, $\mbb{N}_a$ must have at least one negative eigenvalue.

To extend this result, we now show that $\mbb{\tilde N}_a$ has exactly one negative eigenvalue and $|S_a|-1$ positive eigenvalues. This follows directly from an eigenvalue interlacing theorem~\cite{Bun78,Tho76,Wil65}, which states that for symmetric matrices perturbed by a rank-one perturbation as in \Eqnref{rankonepert}, the ordered eigenvalues $K_n$, $K_1\geq K_2\geq\ldots\geq K_{|S_a|}$, of $\mbb{\tilde N}_a$ (and thus of $\mbb{N}_a$) have the property
\algn{
	K_{|S_a|} \leq \tilde \Delta_{|S_a|}\leq K_{|S_a|-1}\leq\tilde\Delta_{|S_a|-1}\ldots\leq \tilde\Delta_{1}\,,
}
where $\tilde\Delta_{k}$ are the ordered diagonal elements of $\mbb{\tilde W}_a$, such that  $\tilde \Delta_1\geq \tilde \Delta_2\geq\ldots\geq \tilde \Delta_{|S_a|}$. In other words, the eigenvalues of $\mbb{\tilde N}_a$ are interlaced with the ordered diagonal elements $\tilde\Delta_k$ of $\mbb{\tilde W}_a$. Since all $\tilde\Delta_k$ are positive for $a<\frac13$, we find that all $K_k$ with $k<|S_a|$ must be positive.

But because the determinant $\det\mbb{\tilde N}_a$ in \Eqnref{Kdet} is always negative, the smallest eigenvalue $K_{|S_a|}$ must be negative. This means that $\mbb{\tilde N}_a$ has exactly one negative eigenvalue and $|S_a|-1$ positive eigenvalues.

We conclude that the complete matrix $\mbb{\tilde N}_a$ (and thus the untransformed $\mbb{N}_a$) has at least $|S_a|-1$ unstable directions, which means that only states with a single active Fourier mode, $|S_a|=1$, can be stable for $a<\frac13$, as stated in the main text.
\section{Thermodynamic observables close to the transition}\seclab{appthermo}
We analyse the behaviour of thermodynamic observables close the synchronisation transition. All observables have the generic, composite form \Eqnref{obsform}. Expanding to second order in $\ve{x}$, we find
\algn{
	\langle \mc{\dot O}\rangle \sim q\mc{\dot O}_{00} +\frac12\sum_{n=0}^{q-1}\left(\mc{\dot O}_{20}x_{n+1}^2+2\mc{\dot O}_{11}x_{n+1}x_n + \mc{\dot O}_{02}x^2_n \right)\,.
}
Applying the Fourier transform we find from \Eqsref{x2ftrans}
\algn{
	\langle \mc{\dot O}\rangle \sim q\mc{\dot O}_{00} +\frac1{2q}\sum_{k'=1}^{q-1}\left(\mc{\dot O}_{20}+\mc{\dot O}_{02}+2\mc{\dot O}_{11}\ee^{-\frac{2\pi k'}q}\right)\hat x_{k'} \hat x_{-k'}\,.
}
$\mc{\dot O}(x,x)$ is linear in $x$. It follows that $\mc{\dot O}_{20}+\mc{\dot O}_{02} = -2\mc{\dot O}_{11}$. Using this, we write
\algn{
	\langle \mc{\dot O}\rangle \sim& q\mc{\dot O}_{00} -\frac{2\mc{\dot O}_{11}}q\sum_{k'=1}^{q-1}\sin^2\left(\frac{\pi k'}q\right)\hat x_{k'} \hat x_{-k'}\,.
}

With the notation in \Eqnref{xzrel} we write the sum as
\algn{
	\sum_{k'=1}^{q-1}\sin^2\left(\frac{\pi k'}q\right)\hat x_{k'} \hat x_{-k'} = \sum_{k'=1}^{\lfloor\frac{q}2\rfloor}\left(1-\frac12\delta_{k'\frac{q}2}\right)\sin^2\left(\frac{\pi k'}q\right)|z_{k'}|^2\,.
}
For $t\to\infty$ and to lowest order in $\Lambda$, we have $|z_{k'}|^2 \sim r^{*2}_{k'}$. This gives
\algn{
	\langle\mc{\dot O}\rangle \sim& \langle\mc{\dot O}\rangle_0 -\frac{4\mc{\dot O}_{11}}q{\ve v}_a\cdot{\ve r}^{*2}_a\,,
}
where $\langle\mc{\dot O}\rangle_0 \equiv q\mc{\dot O}_{00}$.

We now express $\mc{\dot O}_{11}$ in terms of $j_{nm}$. To this end, we note that the composite functions $\mc{\dot O}(y,x)$ in \Eqnref{obsform} can, for all thermodynamic observables $\mc{\dot O}$ be written as
\algn{\eqnlab{ocompgen}
	\mc{\dot O}(y,x) = \dot o(y,x)j(y,x)\,,
}
so that 
\sbeqs{
\algn{
	\mc{\dot O}_{00} &= \dot o_{00}j_{00}\,,\\
	\mc{\dot O}_{11} &= \dot o_{11}j_{00} + \dot o_{10}j_{01} + \dot o_{01}j_{10} + \dot o_{00}j_{11}\,,
}
}
with coefficients $\dot o_{nm} = \partial_y^n\partial_x^m \dot o(y,x)|_{x,y = \frac{1}q}$. $\dot o(y,x)$ has the property
\algn{
	\dot o(y,x) + \dot o(x,y) = \text{const}\,,
}
which implies the general relations
\algn{
	\partial_y\dot o(y,x) = -\partial_{y}\dot o(x,y)\,,\quad \partial_{y}\partial_x \dot o(y,x) = 0\,,
}
We then have for $\dot o_{nm}$,
\algn{
	\dot o_{00} = \text{const}\,,\quad \dot o_{10} = -\dot o_{01}\,,\quad \dot o_{11} = 0\,.
}
We thus find
\algn{
	\mc{\dot O}_{11} = \dot o_{01}(j_{10}-j_{01}) + \dot o_{00}j_{11} \sim \dot o_{00}j_{11}\,,
}
because $j_{10}-j_{01} = \Lambda/2 \ll 1$ close to the synchronisation transition. Putting everything together, we now have
\algn{
	\frac{\langle\mc{\dot O}\rangle-\langle\mc{\dot O}\rangle_0}{|\langle\mc{\dot O}\rangle_0|} \sim&-\frac{4 j_{11}}{q^2j_{00}}{\ve v}_{a}\cdot {\ve r}^{*2}_a = -2\Gamma \lambda  \alpha_a\,,
}
as given in \Eqnref{dO} in the main text.
\section{Phase-space contraction rate close to transition}\seclab{apppscontr}
We derive an expression for the phase-space contraction rate $\mc{L}$ that holds close to the synchronisation transition. We expand the equation of motion \eqnref{eom} to third order in $x_n$ around the decoherent fixed point and substitute the expansion into the definition~\eqnref{pscr} of $\mc{L}$. We find
\begin{multline}
	\mc{L} \sim -q(j_{10}-j_{01}) - \frac1{2}\sum_{n=0}^{q-1}\left[(j_{30}-j_{21})x_{n+1}^2	\right.\\
	\left.+ 2(j_{03}-j_{30})x_{n+1}x_n +  (j_{12}-j_{03})x_n^2 \right]\,.
\end{multline}
Applying the Fourier transforms~\eqnref{x2ftrans} and using the same arguments as in Appendix~\secref{appthermo} leads us to
\algn{
	\mc{L} \sim& -q(j_{10}-j_{01}) + \frac{2(j_{03}-j_{30})}{q}\sum_{k'=1}^{q-1}\sin^2\left(\frac{\pi k'}q\right){\hat x}_{k'}{\hat x}_{-k'}\,,\nn\\
	\sim& -\frac{q\Lambda}{2} + qA{\ve v}_{a}\cdot {\ve r}^{*2}_a = -\frac{q\Lambda}2 + q\Lambda \alpha_a\,.
}
With $\mc{L}_0=-q\Lambda/2$, we readily obtain \Eqnref{dL} in the main text.
\end{document}